# Photoluminescent Tetragonal Tb-doped $Pb_2P_2O_7$

*Yong Liu, Wenhua Bi, Alla Arakcheeva and Arnaud Magrez*

Crystal Growth Facility, Institute of Physics, Ecole Polytechnique Fédérale de Lausanne, CH-1015 Lausanne

# Abstract

In this study, we report the synthesis and characterization of a novel tetragonal polymorph of Tb-doped $Pb_2P_2O_7$. Single-crystal X-ray diffraction confirms the stabilization of the $P4_1$ and $P4_3$ enantiomorphs at room temperature due to the incorporation of $Tb^{3+}$ ions. Optical investigations reveal green photoluminescence from the characteristic $^5D_4 \rightarrow {}^7F_j$ (J = 1–5) transitions of $Tb^{3+}$, with each emission split due to the crystal field effect, indicating the presence of $Tb^{3+}$ in multiple coordination environments. The power dependence of the PL intensity follows a linear power-law behavior, suggesting a one-photon excitation process. Temperature-dependent PL measurements show an initial increase in intensity up to 125°C, attributed to energy transfer from structural defects, followed by thermal quenching above this temperature. Structural stability at elevated temperatures is confirmed via high-temperature X-ray diffraction (XRD), showing no phase transitions before melting at approximately 800°C. These findings highlight the potential of tetragonal Tb-doped $Pb_2P_2O_7$ as a new class of photoluminescent material.

# Introduction

Pyro inorganic compounds, with composition $R_2M_2O_7$, exhibit a structure consisting of isolated edge-sharing $(M_2O_7)^{-2n}$ double tetrahedra or octahedra and $R^{+n}$ cation. This class of compounds can be divided in three groups: Thorthievite, Dichromates and Atopites. In Atopites, the coordination number of M is equal to 6 while in Thorthievite and dichromates the coordination number of M is equal to 4. The Thorthievite and Dichromates differ from the configuration of the $(M_2O_7)^{-2n}$ in which tetrahedra can be respectively staggered or eclipsed depending of the size of the $R^{+n}$ cation. Pyrophosphates belong to the Dichromate group if R > 0.8 Å and to Thortievites otherwise.[1]

According to ICSD,[2] COD,[3] and PDF5 databases,[4] there are 16 known pyrophosphate compounds with R= Mg, Ca, Sr, Ba, Cr, Mn, Fe, Co, Ni, Pd, Cu, Zn, Cd, Hg, Sn and Pb. All pyrophosphates crystallize in monoclinic or triclinic symmetry, except $Zn_2P_2O_7$, $Sr_2P_2O_7$ and $Ba_2P_2O_7$ having orthorhombic structure. Highest symmetries are observed in $Ca_2P_2O_7$ and $Ba_2P_2O_7$ crystallizing in tetragonal $P4_1$ and hexagonal $P$-62$m$ space groups respectively. $Pb_2P_2O_7$ is a $n$-type semiconductor exhibiting a low surface tension and strong $\sigma$-acceptor behavior which facilitates the surface adsorption of polybutadiene epoxide (PBDE). [5] Thin films of PBDE on inorganic surfaces may represent a new class of reactive coatings to be used as a polymer bonding/coupling agent in various solid/solid interfaces. [6] Lead pyrophosphate was also found to form during drinking water treatment when phosphate additives are used to reduce lead level. [7] $Pb_2P_2O_7$ can be grown either as crystals or as glass. [8-10] These materials have high optical quality, [11] chemical stability and radiation hardness. Thanks to these properties, $Pb_2P_2O_7$ was theoretically considered as potential new scintillator. [12]

Lead pyrophosphate crystals were first grown by the Czochralski technique. Feed material was prepared by reaction of lead carbonate and ammonium dihydrogen phosphate. The crystal structure is refined in a triclinic symmetry with $P$-1 space group. Refined lattice parameters are $a$ = 6.9627 Å, $b$ = 6.9754 Å, $c$ = 12.764 Å, $\alpha$= 96.78°, $\beta$= 91.16° and $\gamma$= 89.68°. [13] Similar results are obtained with crystals obtained by reaction of $K_2Pb[P_4O_{12}]$ with moisture, [14] or by decomposition of $PbHPO_4$. [15] A detailed structural analysis confirms $Pb_2P_2O_7$ to belong to the dichromate group of pyrophosphates with an eclipsing dihedral angle as high as 13°. [16] At 670°C, $Pb_2P_2O_7$ was shown to exhibit a phase transition to a monoclinic structure with $P2_1/n$ space group. The temperature phase transition is reversible. [17] The high temperature polymorph can be stabilized at room temperature by partial substitution of lead by strontium. [18] The high temperature structure of $Pb_2P_2O_7$ is isomorphic to $\alpha$-$Ca_2P_2O_7$. Below approximately 1170°C, $Ca_2P_2O_7$ exhibits a structure with $P4_1$ space group. [19] This tetragonal $\beta$-$Ca_2P_2O_7$ structure was later revisited by Boudin et al. [20] and more accurate atomic parameters are available. The new description of the pyrophosphate groups in $Ca_2P_2O_7$ reveals the P-O bonds involving the bridging O atoms are longer than the non-bridging P-O bonds of the tetrahedra. Ca containing polyhedra are reported with greater accuracy as bicapped trigonal prism, tricapped trigonal prism, and distorted pentagonal bipyramid. [20]

According to the databases, [2-4] the $\beta$-$Ca_2P_2O_7$ class of materials is limited to lanthanide pyrosilicates $Ln_2Si_2O_7$ (Ln from La to Lu), $Sr_2V_2O_7$ and $Sr_2As_2O_7$. Both $P4_1$ and $P4_3$ enanthiomorphs of strontium pyroarsenates have been produced and high-quality refined structures are available. [21,22] Although the $P4_1$ tetragonal crystals of $Sr_2V_2O_7$ were grown in $NaVO_3$ flux, [1] the properties of $Sr_2V_2O_7$ were solely studied in the triclinic polymorph. In addition to the dielectric properties, [23] highly toxic hexavalent chromium is efficiently photo-reduced by strontium pyrovanadates when blended with carbon nitride. [24] When doped with $Eu^{3+}$ or $Dy^{3+}$, monoclinic strontium pyrovanadate phosphors emit white light and were applied in LEDs. [25-27] Lanthanide pyrosilicates ($Ln_2Si_2O_7$) exhibit up to 7 polymorphs, [28] and are also intensively studied for their optical properties but, like $Sr_2V_2O_7$, studies of lanthanide pyrosilicates properties and their applications are limited to the monoclinic polymorph. Pyrophosphate compounds with a tetragonal symmetry remain relatively underexplored. The $P4_1$ and $P4_3$ space groups are non-centrosymmetric and materials crystallizing in such symmetries may potentially exhibit piezoelectric, ferroelectric as well as second-order nonlinear optical behavior. [29] Furthermore, the incorporation of photoluminescent terbium, combined with the high thermal stability characteristic of pyrophosphate compounds, suggests potential applications in high-power LED lighting systems (e.g., spotlights, headlights, and industrial lighting), optical temperature sensing for remote thermography, signaling devices for extreme environments such as aerospace and space applications, and high-temperature imaging in metallurgical and chemical processes.

This work demonstrates that small amounts of $Tb^{3+}$ doping are sufficient to induce a symmetry change in $Pb_2P_2O_7$, from the centrosymmetric triclinic $P$-1 structure to the non-centrosymmetric tetragonal $P4_1$ or $P4_3$ space groups. The resulting Tb-doped $Pb_2P_2O_7$ crystals adopt a $\beta$-$Ca_2P_2O_7$-type structure. The non-centrosymmetric nature of this phase is confirmed by single-crystal X-ray diffraction. In addition, preliminary photoluminescence results are presented, highlighting the optical response of the material.

# Experimental

**Powder synthesis**

Monoclinic $Pb_2P_2O_7$ powder was synthesized by a solid-state reaction method. The reaction was carried out with lead carbonate, $PbCO_3$, and ammonium dihydrogen phosphate, $(NH_4)H_2PO_4$, according to following equation: [14]

$$2PbCO_{3(s)} + 2(NH_4)H_2PO_{4(s)} \rightarrow Pb_2P_2O_{7(s)} + 2NH_{3(g)}\uparrow + 3H_2O_{(g)}\uparrow + 2CO_{2(g)}\uparrow \qquad \text{Eq. 1}$$

$PbCO_3$ powder and $(NH_4)H_2PO_4$ grains were ground into fine powder in an agate mortar. The powder was saved in a platinum crucible and loaded into a box furnace. The raw materials were heated at 300°C for a period of 12 h and then at 700°C for extra 12 h. X-ray diffraction measurement reveals the formation of $Pb_2P_2O_7$ phase with triclinic structure, as shown in Figure 1 and Figure S4.

Tetragonal Tb-doped $Pb_2P_2O_7$ powders were prepared by grinding tetragonal Tb-doped $Pb_2P_2O_7$ single crystals. The description of Tb-doped $Pb_2P_2O_7$ single crystal growth can be found in the next paragraph. After the growth, the crystals were separated from the flux by washing with diluted $HNO_3$ solution. Attempts to synthesize Tb-doped $Pb_2P_2O_7$ using $Pb_2P_2O_7$ and $Tb_4O_7$, instead of $TbPO_4$ like during the crystal growth were unsuccessful. These results suggest that $TbPO_4$ is a necessary precursor for effective incorporation of $Tb^{3+}$ in $Pb_2P_2O_7$ and for stabilizing the tetragonal phase.

**Single crystal growth**

Lead pyrophosphate has a low temperature melting point of about 800°C. It melts congruently with low vapor pressure. Therefore, it is a frequently used flux for the growth of single crystals. Tetragonal Tb-doped $Pb_2P_2O_7$ single crystals were synthesized during the growth of $TbPO_4$ single crystals using $Pb_2P_2O_7$ flux. [30,31] In our study, the $TbPO_4$ and $Pb_2P_2O_7$ powders were mixed at a molar ratio of $TbPO_4 : Pb_2P_2O_7 = 1:19$ and loaded into platinum crucible. The mixture was heated up to 1300°C and hold for 12 h. After the dwell at high temperature, the furnace was switched off and naturally cooled down to room temperature. Brown tetragonal Tb-doped $Pb_2P_2O_7$ single crystals can be readily cleaved from the solidified melt or released by dissolving the flux with diluted nitric acid. $Pb_2P_2O_7$ single crystals with the triclinic structure were obtained in a Pt crucible by heating $Pb_2P_2O_7$ powder up to 1000°C and cooling down to 750 °C. [17]

**XRF analysis**

X-Ray fluorescence measurements were performed using an OrbisPC Micro EDXRF analyzer system, equipped with a Rh micro-focus X-ray tube (50kV and 1mA), a 30$\mu$m poly-capillary X-Ray optics, and an Apollo XRF-ML50 silicon drift detector with an energy resolution lower than 135eV. All the XRF Measurements reported here, were performed under vacuum conditions.

**Single-Crystal X-Ray diffraction experiments**

Crystals of triclinic and tetragonal Tb-doped $Pb_2P_2O_7$ with suitable size was selected and mounted on goniometer head with a cryo-loop. Frames were collected at 100 K on a Rigaku Synergy-I XtaLAB X-ray diffractometer, equipped with a Mo micro-focusing source ($\lambda K_\alpha$ = 0.71073 Å) and a HyPix-3000 Hybrid Pixel Array detector (Bantam). The temperature was controlled by a Cryostream 800 from Oxford Cryosystems Ltd. *CrysAlisPro*,[32] and *OLEX²* software, [33] were used for data reduction and structural refinements, respectively. Structure solutions were obtained with ShelXT program. [34] Other experimental details are listed in **Table S1 and S2**, for different enantiomers respectively and polymorphs. Further details are available as a CIF file with CSD numbers 2502760 and 2502765 for the $P4_1$ and $P4_3$ enantiomorphs respectively. VESTA was used for crystal structure plotting. [35]

**Polycrystalline X-Ray diffraction experiments**

Powders were transferred on zero-background sample holders and mounted on a Panalytical Empyrean X-ray polycrystalline diffractometer, equipped with long-focused Cu X-ray tube ($\lambda K_{\alpha}$ = 1.5418 Å), and PIXcel 1D X-ray detector. The patterns were collected in continuous mode between 10° and 80° in $2\theta$, with the step-size of 0.02626°.

Non-ambient PXRD scans were performed on a PHILIPS X'Pert MPD PRO diffractometer, equipped with long-focused Cu X-ray tube ($\lambda K_{\alpha}$ = 1.5418 Å), and X'Celerator 1D X-ray detector. The patterns were collected in continuous mode between 10° and 80° in $2\theta$, with the step-size of 0.0167°. The temperature was controlled by an Anton Paar HTK1200N sample heating chamber, with capillary extent device for spinning the fine powder in quartz capillary during the measurements. The pattern reduction, as well as the following search-match, were performed with HighScore plus v4.9, [36] and PDF5 v2021. [4]

**Photoluminescence measurements**

Photoluminescence spectroscopy was taken using inVia™ Raman microscope (Renishaw, Gloucestershire, UK) with excitation wavelengths of 488 nm.

# Results and Discussion

Following the growth process, $TbPO_4$ crystals are embedded within a matrix composed of brownish $Pb_2P_2O_7$ flux. The dissolution of this flux using diluted nitric acid results in the collection of octahedral brown crystals, alongside $TbPO_4$ needles, as illustrated in Figure 1a and Figure S1. These crystals exhibit noteworthy dimensions, reaching up to one centimeter. Analysis of the X-ray diffraction (XRD) pattern obtained from a selection of crystals ground into powder indicates a structural distinction between the brown crystals and the previously reported triclinic $Pb_2P_2O_7$ (Figure 1c). X-ray fluorescence (XRF) analysis corroborates a Pb/P ratio in close to 1 similarly to the stoichiometry of $Pb_2P_2O_7$ and verifies the presence of Tb (Figure 1d). Importantly, XRF measurements were conducted on the interior of the crystals, after cutting and polishing procedures. This spatial analysis confirms the internal presence of Tb within the crystals. Further investigation of ten distinct brown crystals reveals varying concentrations of Tb relative to Pb, ranging from 1.1% to 3.7% (Figure S2).

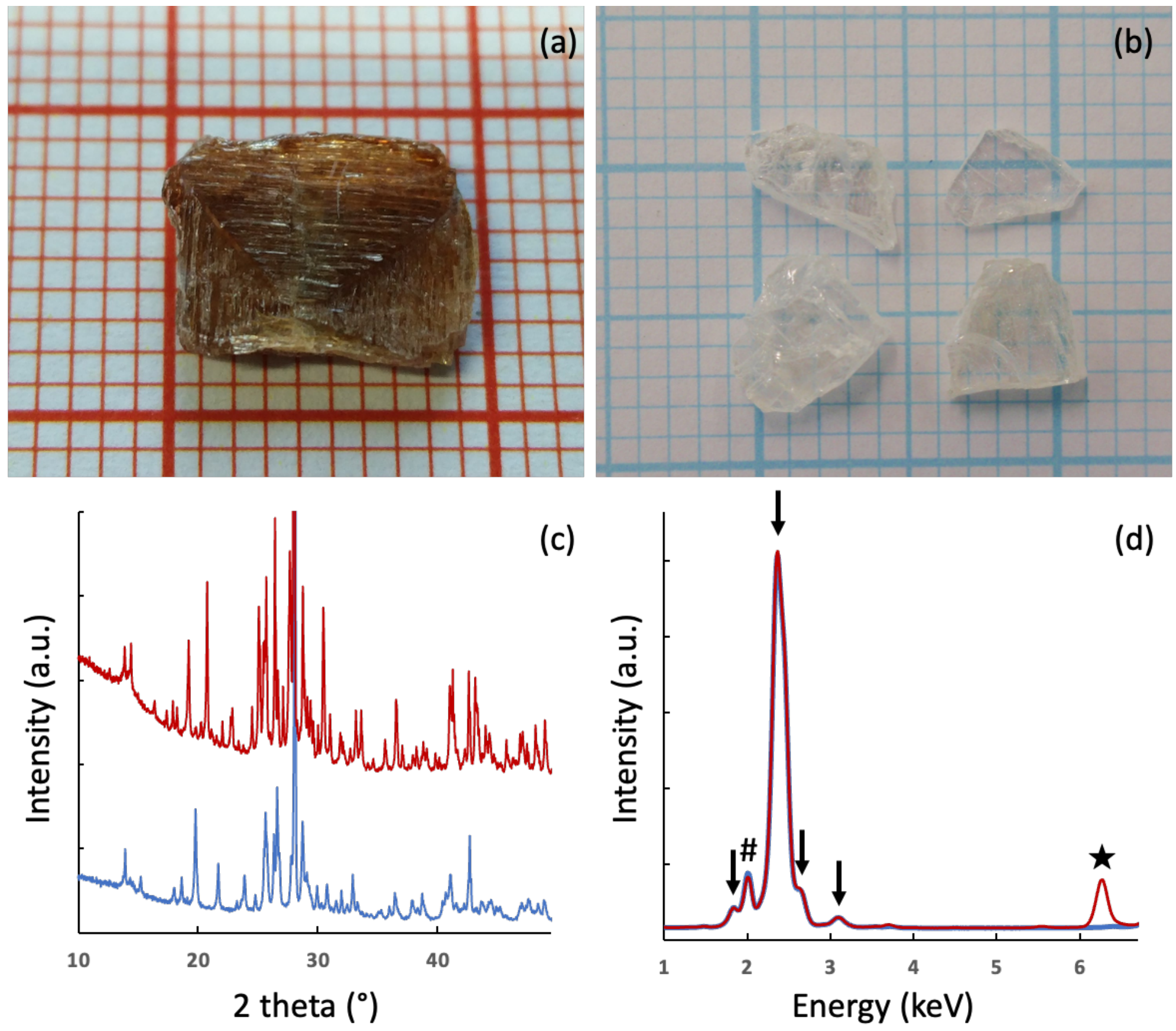

**Figure 1**: Optical pictures of the tetragonal Tb-doped $Pb_2P_2O_7$ (a) and triclinic $Pb_2P_2O_7$ (b) crystals. The corresponding powder XRD patterns (c) as well as the XRF spectra (d) are shown. Red and blue patterns correspond to the tetragonal Tb-doped $Pb_2P_2O_7$ and triclinic $Pb_2P_2O_7$ respectively. The comparison of the experimental XRD pattern with the simulated pattern is provided in Figure S4. In the XRF patterns, lead, phosphorus and terbium are indicated by arrows, hashtag and star respectively.

**Structure description**

Figure 2 displays the experimental data from single crystal XRD conducted on a high-quality brown crystal. The *hk*0 cross section of the reciprocal space reconstruction allows for the unequivocal determination of the cell parameters $a = b = 6.96$ Å (Figure 2a). Additionally, the $h0l$ cross section, equivalent to the $0kl$ cross section (not shown), indicates that the unit cell is $P$-centered tetragonal with $c$ = 25.36 Å (Figure 2b). The observed systematic extinction of reflections ($00l$: $l = 4n$) signifies the presence of a $4_1$ or $4_3$ screw axis along the $c$-axis, leading to the assignment of the space groups $P4_1$ and $P4_3$, similar to $\beta$-$Ca_2P_2O_7$. Through the combined XRF analysis and single crystal XRD data, it is established that the brown crystals are Tb-doped $Pb_2P_2O_7$ isomorphic to $\beta$-$Ca_2P_2O_7$. Room temperature structural refinement was carried out in the $P4_1$ and $P4_3$ space groups, as detailed in Table S1, S3 and S4. The crystals exhibit variability, with both pure $P4_1$ and $P4_3$ enantiomorphs present, along with racemic ones showing a $P4_1$ fraction ranging from 15 to 65%. Complementary crystallographic data obtained from an undoped triclinic $Pb_2P_2O_7$ crystal, similar to those in Figure 1b, are provided in the supporting information Table S2 and S5.

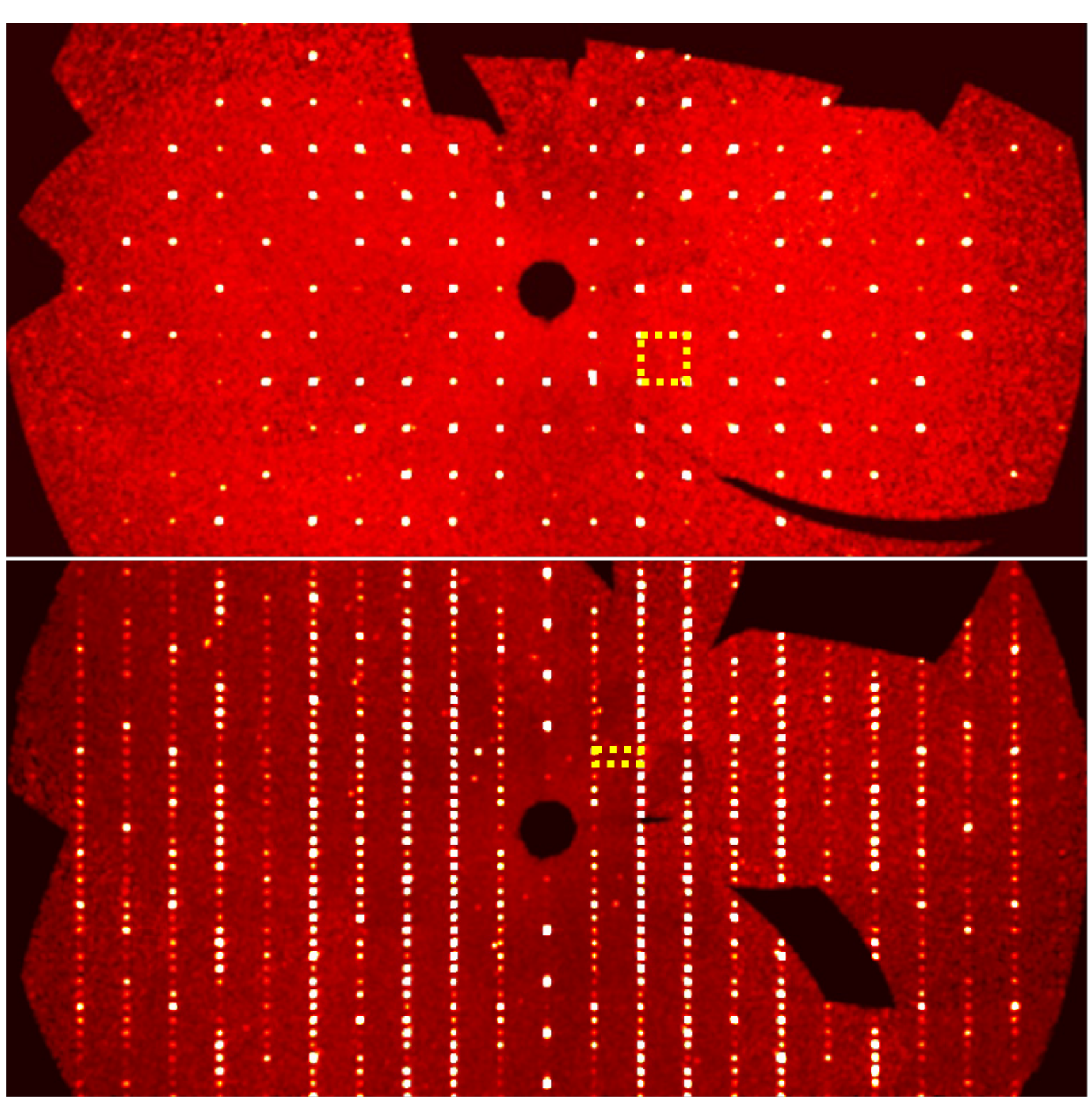

**Figure 2:** Representative cross sections of the reciprocal space reconstruction. The $hk0$ cross section confirms the tetragonal symmetry of the lattice with $a$ = $b$ = 0.696 nm. No specific

extinction rule can be observed (up). The *h*0*l* similar to the 0*kl* cross section confirms the *P*-centered tetragonal lattice with $c$ = 2.536 nm. 00*l* reflections are visible if $l = 4n$ which is consistent with a $4_1$ (or $4_3$) screw axis along the *c*-axis (down). The reciprocal tetragonal unit cell is marked in yellow.

Isomorphous of $\beta$−$Ca_2P_2O_7$, tetragonal Tb-doped $Pb_2P_2O_7$ structure is delineated as a layered arrangement, visually represented in Figure 3. Each unit cell comprises four layers, with four Pb ions situated in each layer. Within the layers, the shortest distance between pyrophosphate groups at 100 K is measured at 4.012 Å. The layers are bonded by pyrophosphate groups through $Pb^{2+}$ ions. The relationship between the layers is governed by the four-fold screw axis along the *c*-axis. The distance between pyrophosphate groups between two successive layers is 4.881 Å. For comparison, the pyrophosphate distance within the layer is identical in the triclinic and tetragonal polymorph while the distance between layers is measured shorter at 4.736 Å in the triclinic $Pb_2P_2O_7$.

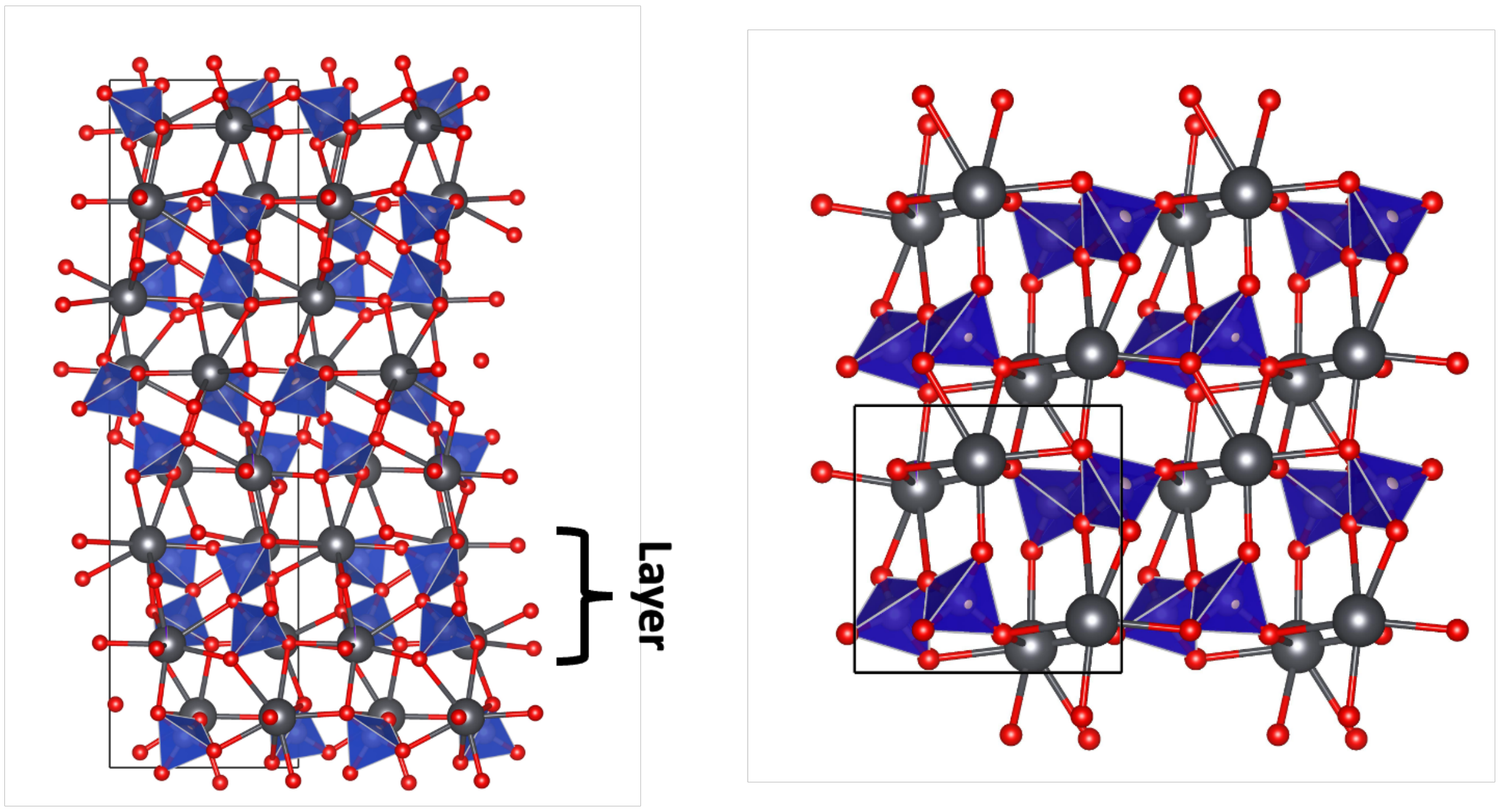

**Figure 3**: Tetragonal Tb-doped $Pb_2P_2O_7$ structure represented perpendicular to the (100) direction (left). In plane view of the layer composing the structure showing the pyrophosphates groups are organized in a distorted square lattice within the layer (right).

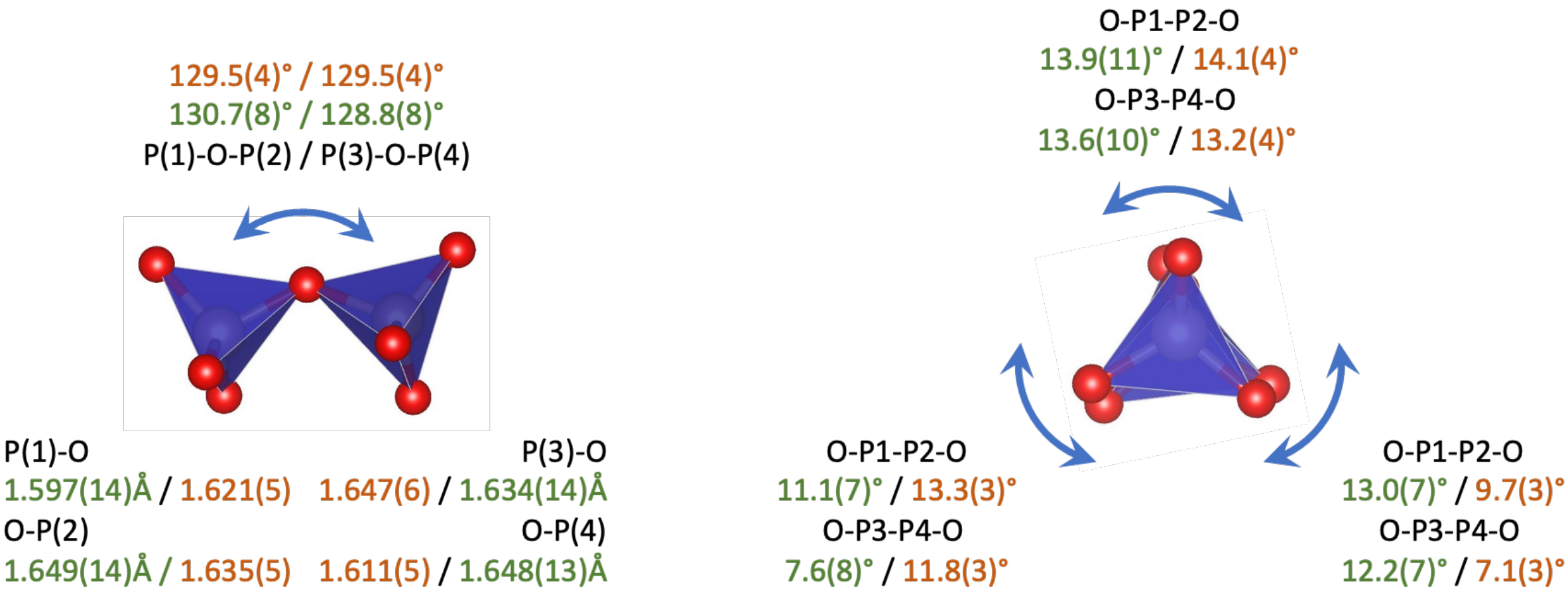

**Figure 4**: Bridging (left) and eclipsing (right) angles as well as interatomic distances of the pyrophosphate groups tetragonal $P4_1$ (green) and tetragonal $P4_3$ (brown) structures at 100 K. The characteristics of the pyrophosphate groups in the triclinic structure are given in Figure S5.

In both polymorphs belonging to the dichromates, the anticipated eclipsing of pyrophosphate groups is observed. Figure 4 presents the variation of eclipsing angles, ranging from 7.8° to 15.2° in the *P*-1 space group and from 7.6° to 14.1° in the tetragonal space groups. Bridging angles between the two $PO_4$ tetrahedra are around 130°, while the bridging P-O-P distances are symmetric and remain approximately at 1.62 Å for both structures. In summary, the pyrophosphate groups exhibit similarities in the triclinic $Pb_2P_2O_7$ and in the tetragonal Tb-doped $Pb_2P_2O_7$.

Because of its lone pair, the coordination of lead in the two $Pb_2P_2O_7$ polymorphs is complicated to determine. As reported by Mullica et al. for the triclinic polymorph, [17] Pb(1) exhibits coordination to eight oxygen atoms arranged in a pattern resembling a bi-augmented trigonal prism. Both Pb(2) and Pb(3) are nine-coordinated, adopting a geometric configuration resembling a distorted mono-capped square antiprism. Pb(4) is bonded to eight oxygen atoms, characterized as a severely distorted square antiprism. Similarly to $\beta$−$Ca_2P_2O_7$,[20] Pb(1) and Pb(3) are eight coordinated in bicapped trigonal prisms. Pb(2) is surrounded by nine oxygen atoms which delimit a tricapped trigonal prism. Pb(4) is linked to seven O atoms which form a distorted pentagonal bipyramid. Selected interatomic Pb-O distances for both tetragonal and triclinic structures are available in Table S6.

The distribution of $Tb^{3+}$ ions over the crystallographic sites was refined using single-crystal XRD. As shown in Table S2 and S3, $Tb^{3+}$ is detected on two of the four $Pb^{2+}$ sites. These results suggest no strong site preference, although the absence of detectable $Tb^{3+}$ on the two other sites in the two crystals analyzed may be due to a concentration below the detection threshold.

**Behavior at high temperature**

As can be seen in the DSC curves in the Figure S3, both triclinic and tetragonal polymorphs of $Pb_2P_2O_7$ melts at approximately 800°C. No additional peaks prior to melting are observed suggesting the absence of phase transition for both polymorphs. It is confirmed by high temperature XRD performed up to 800°C (Figure 5a and 5d).

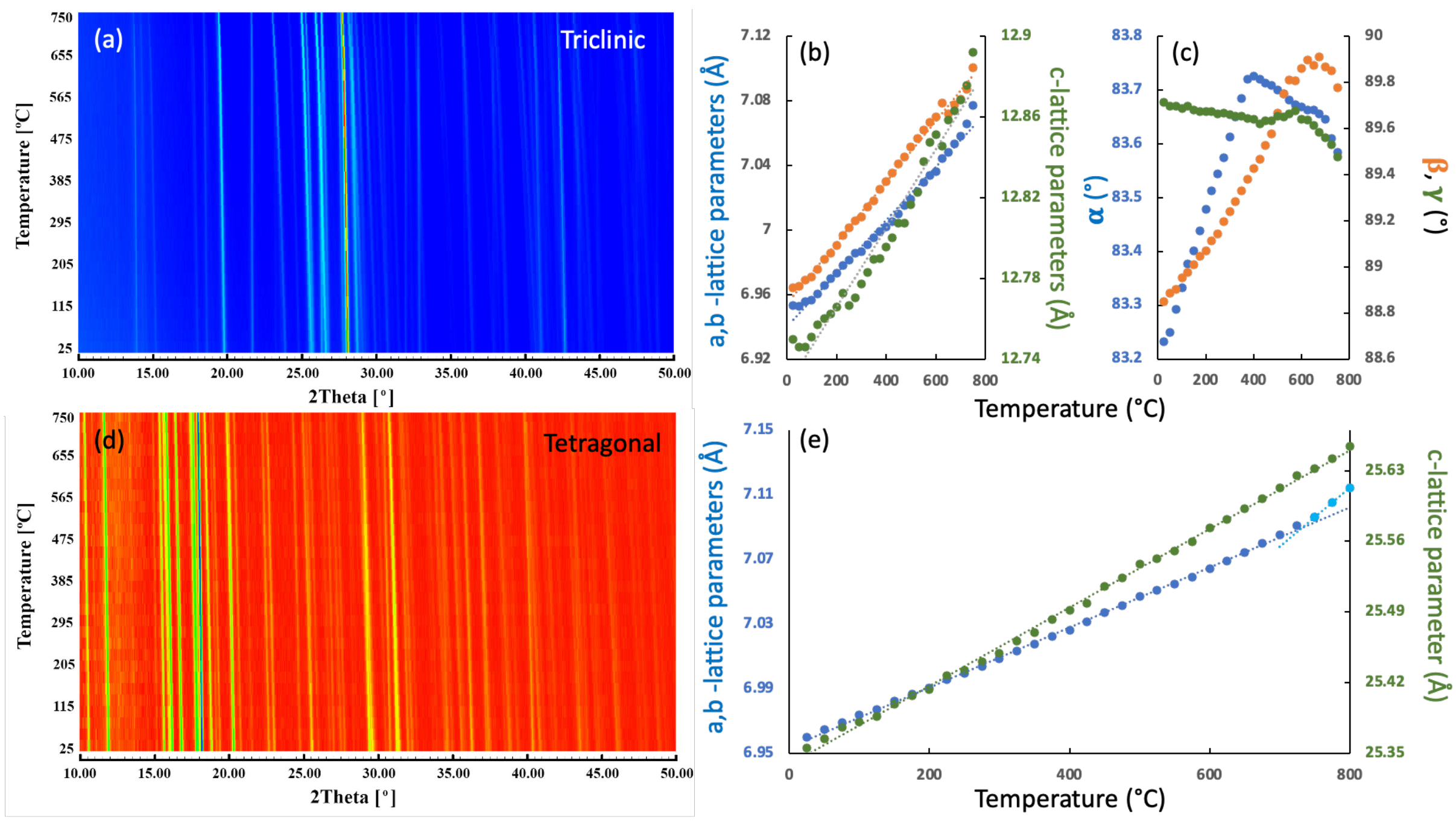

**Figure 5**: High Temperature powder X-Ray diffraction. Contour plot showing the evolution of the XRD patterns with temperature for the triclinic $Pb_2P_2O_7$ (a). (b) and (c) shows the temperature evolution of the lattice parameters for the triclinic $Pb_2P_2O_7$ structure. Contour plot showing the evolution of the XRD patterns with temperature for the tetragonal Tb-doped $Pb_2P_2O_7$. (e) shows the temperature evolution of the lattice parameters for the tetragonal Tb-doped $Pb_2P_2O_7$. Above 750°C the *a*- and *b*- lattice parameters evolution is more pronounced. Higher thermal expansion coefficient is obtained.

Figure 5b, 5c, and 5d illustrate the changes in lattice parameters with temperature. For both polymorphs, parameters *a*, *b*, and *c* increase nearly linearly up to their melting points. However, in the triclinic polymorph, distinct discontinuities in the evolution of lattice angles are observed between 400°C and 600°C. The triclinic polymorph exhibits anisotropic thermal expansion, with the expansion coefficient of the *a*-parameter ($2.3 \times 10^{-4}$ $K^{-1}$) being significantly higher than those of the *b*- and *c*-parameters ($2.3 \times 10^{-5}$ $K^{-1}$ and 2.8 $10^{-5}$ $K^{-1}$, respectively). In contrast, the tetragonal polymorph shows isotropic thermal expansion, with expansion coefficients of $2.7 \times 10^{-5}$ $K^{-1}$ for the *a* (= *b*) parameter and $1.5 \times 10^{-5}$ $K^{-1}$ for the *c*-parameter. Notably, the expansion of the *a* (= *b*) parameter in the tetragonal polymorph increases near the melting point, with the coefficient rising to $5.1 \times 10^{-5}$ $K^{-1}$.

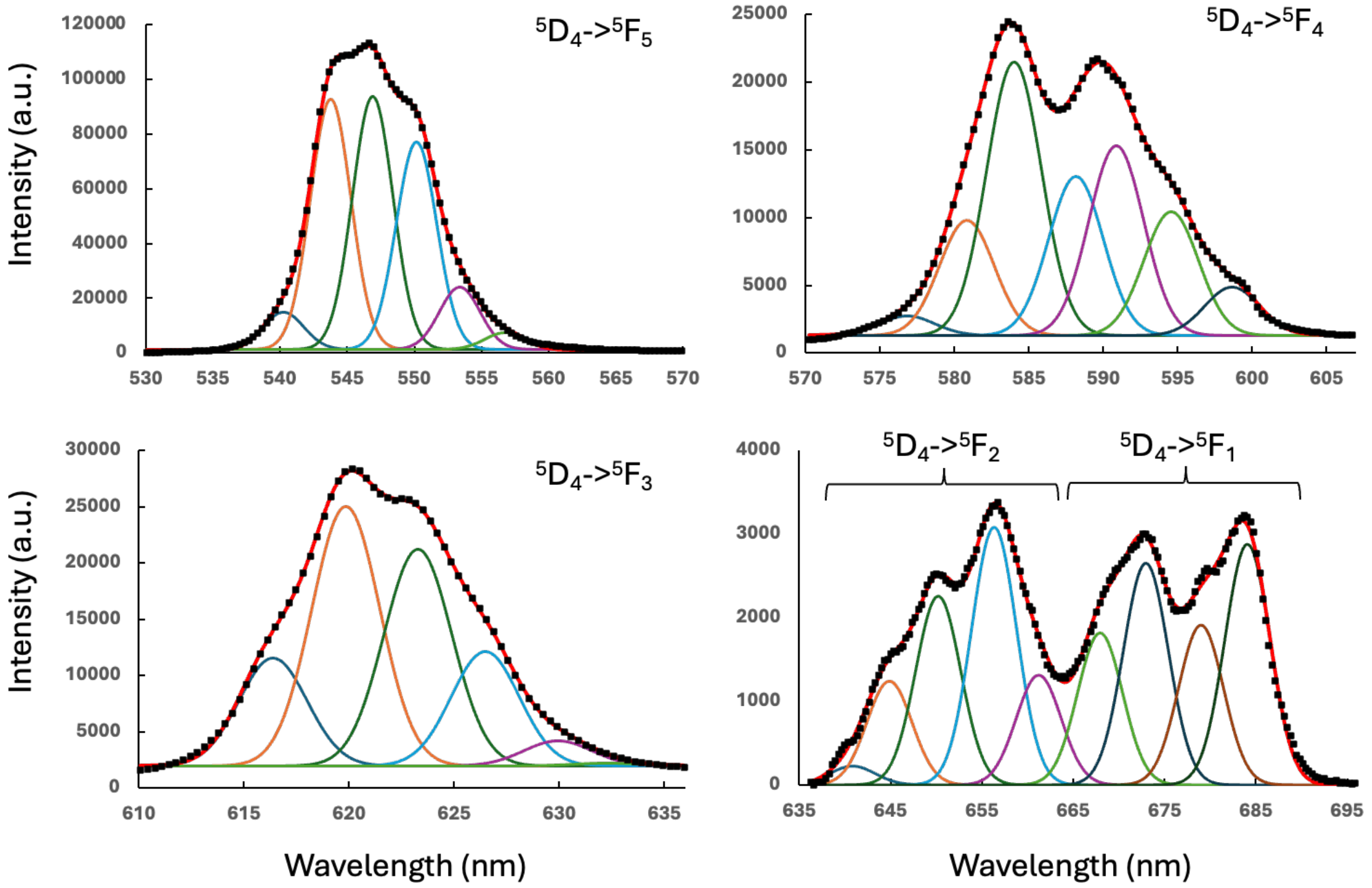

**Figure 6**: Photoluminescence spectra of tetragonal $Tb^{3+}$ doped $Pb_2P_2O_7$. The continuous red curve is the cumulative curve of the individual peaks while the black squares correspond to the experimental data.

**Preliminary photoluminescence measurements**

Photoluminescence (PL) analysis was conducted on a single crystal of tetragonal Tb-doped $Pb_2P_2O_7$ containing 1.1% $Tb^{3+}$, as verified by X-ray fluorescence (XRF). The observed PL originates from the intrinsic 4f-4f electronic transitions of $Tb^{3+}$, resulting in a bright green emission. The PL spectrum, shown in Figure 6, displays five distinct transitions corresponding to $^5D_4$ –> $^7F_j$ (j = 1-5), occurring at approximately 545 nm, 590 nm, 620 nm, 655 nm, 680 nm.

Due to the crystal field effect, each $^5D_4$ –> $^7F_j$ transition is split into multiple components, indicating that $Tb^{3+}$ ions occupy different crystallographic sites with different local coordination environments. The presence or absence of lead vacancies nearby could explain the presence of more than 4 components in the emission lines.

The spatial distribution of $Tb^{3+}$ ions was estimated using the formula:

$$d_{Tb} = 2\left(\sqrt[3]{\frac{3V}{4\pi CN}}\right)$$

where $V$ represents the unit cell volume of $Pb_2P_2O_7$ (1228.5 Å$^3$), $N$ is the number of cationic sites available for $Tb^{3+}$ substitution (4 sites per unit cell), and $C$ denotes the Tb concentration. Based on this calculation, the average separation between $Tb^{3+}$ ions is found to be approximately 38 Å. Although further studies on the PL lifetime are required, this separation suggests that cross-relaxation processes between $Tb^{3+}$ ions likely occur through the electric multipole interaction mechanism rather than the exchange interaction mechanism. [37]

As illustrated in Figure 7a, the PL intensity as a function of excitation power follows a power-law dependence, with a power-law exponent close to unity for all five observed emissions. This near-linear behavior suggests that the excitation mechanism involves a one-photon absorption process and remains valid up to the highest available laser excitation power.
The temperature-dependent evolution of PL intensity, depicted in Figure 7b, reveals an initial increase from room temperature to 125°C for all emission lines. This enhancement may be attributed to an improved energy transfer from structural defects, i.e. lead vacancies, compensating the $Tb^{3+}/Pb^{2+}$ charge imbalance. The maximum PL intensity is recorded at 125°C, after which a gradual decline is observed due to thermal quenching. At 600°C, the PL intensity has decreased by approximately 50–60% compared to its peak value at 125°C.

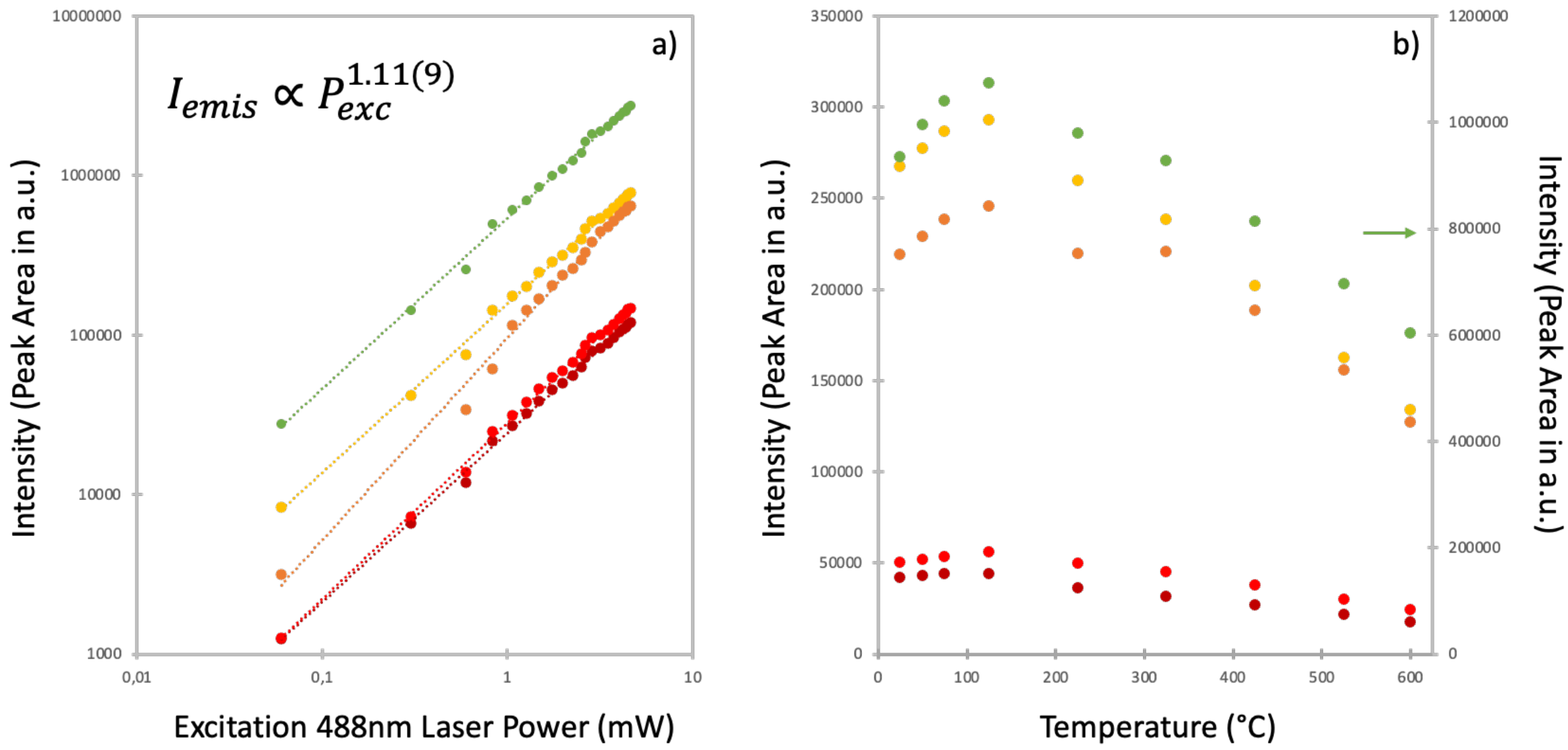

**Figure 7**: (a) Evolution of the photoluminescence intensity as a function of excitation laser power. The power law exponent is calculated as the average of the power law exponents of the five emissions. (b) Temperature dependence of the Photoluminescence intensity of the five emissions of $Tb^{3+}$ in tetragonal $Pb_2P_2O_7$. For clarity, the $^5D_4 \rightarrow {}^7F_5$ emission is plotted using the right ordinate axis while the four $^5D_4 \rightarrow {}^7F_{4-1}$ are plotted using the left ordinate axis.

**Conclusion**:

We have successfully synthesized and characterized a tetragonal polymorph of Tb-doped $Pb_2P_2O_7$, stabilized by terbium doping. Structural refinement reveals that the material crystallizes in the $P4_1/P4_3$ non-centrosymmetric space groups, similar to $\beta$-$Ca_2P_2O_7$, with minimal structural changes between 100 K and 300 K. Photoluminescence studies confirm that $Tb^{3+}$ is incorporated into multiple crystallographic sites, leading to well-defined split emission lines due to the crystal field effect. The excitation power dependence of the PL intensity follows a linear power law, indicating a one-photon absorption process. Thermal evolution studies reveal an initial enhancement of PL intensity up to 125°C, likely due to defect-assisted energy transfer, followed by a gradual quenching at higher temperatures, in line with non-radiative relaxation mechanisms. High-temperature XRD confirms the structural stability of both tetragonal and triclinic $Pb_2P_2O_7$ polymorphs up to their melting point of ~800°C.

Future work should focus on the synthesis of other lanthanide-doped $Pb_2P_2O_7$ systems to investigate the influence of ionic radius, charge and vacancies on the stabilization mechanism of this novel polymorph. Complementary theoretical calculations would provide valuable insight into these effects.

**Acknowledgment**:

Arnaud Magrez acknowledges Raphaël Butté for the fruitful discussions on the photoluminescence of data.

**Supporting information**

Optical microscopy image of $TbPO_4$ needles embedded in Tb doped $Pb_2P_2O_7$ flux. XRF spectra of Tb-doped $Pb_2P_2O_7$ crystals. DSC and XRD pattern profil fitting of $Pb_2P_2O_7$ and Tb-doped $Pb_2P_2O_7$. Characteristics of the pyrophosphate groups in triclinic $Pb_2P_2O_7$. Experimental details of single crystal XRD and interatomic distances in $Pb_2P_2O_7$ and Tb-doped $Pb_2P_2O_7$.

**References:**

(1) Baglio, J. A.; Dann, J. N. The Crystal Structure of β-Strontium Pyrovanadate. *J. Solid State Chem.* **1972**, *4*, 87–93. https://doi.org/10.1016/0022-4596(72)90136-3.

(2) National Institute of Standards and Technology. *NIST Inorganic Crystal Structure Database*; NIST Standard Reference Database Number 3; Gaithersburg, MD. https://doi.org/10.18434/M32147.

(3) Gražulis, S.; Daškevič, A.; Merkys, A.; Chateigner, D.; Lutterotti, L.; Quirós, M.; Serebryanaya, N. R.; Moeck, P.; Downs, R. T.; Le Bail, A. Crystallography Open Database (COD): An Open-Access Collection of Crystal Structures and Platform for World-Wide Collaboration. *Nucleic Acids Res.* **2012**, *40* (D1), D420–D427. https://doi.org/10.1093/nar/gkr900.

(4) Gates-Rector, S.; Blanton, T. The Powder Diffraction File: A Quality Materials Characterization Database. *Powder Diffr.* **2019**, *34* (4), 352–360. https://doi.org/10.1017/S0885715619000812.

(5) Koizumi, K.; Hashizume, D.; Umeda, J.; Kimura, M.; Yokota, T.; Ito, M.; Iguchi, A.; Nakamura, S. A Comparison of Geometries and Electronic Structure of Plumbogummite T ($PbAl_3P_2O_{14}H_6$), $Pb_2P_4O_{12}$, and $Pb_2P_2O_7$. *Chem. Phys. Lett.* **2020**, *756*, 137800. https://doi.org/10.1016/j.cplett.2020.137800.

(6) Seifert, G.; Hortschansky, P.; Heublein, E. L.; Heublein, G. Investigations of Adsorption Behaviour of Polybutadiene Epoxide on Solid Surfaces. *Colloids Surf.* **1991**, *58*, 33–46. https://doi.org/10.1016/0166-6622(91)80196-U.

(7) Lytle, D. A.; White, C.; Schock, M. R. Synthesis of Lead Pyrophosphate, $Pb_2P_2O_7$, in Water. *Microsc. Microanal.* **2008**, *14*, 335–341. https://doi.org/10.1017/S1431927608080689.

(8) Sreeram, A. N.; Hobbs, L. W. Characterization of Metamict and Glassy Lead Phosphates. *MRS Proc.* **1993**, *321*, 25–31. https://doi.org/10.1557/PROC-321-25.

(9) Sales, B. C.; Ramey, J. O.; Boatner, L. A. Structural Alterations in the Amorphous-to-Crystalline Transformation of Lead Pyrophosphate. *Phys. Rev. Lett.* **1987**, *59* (15), 1718–1721. https://doi.org/10.1103/PhysRevLett.59.1718.

(10) Ananthraj, S.; Rao, K. J. Formation of Lead Pyrophosphate Glass and the Role of Anion Disproportionation. *Proc. Indian Acad. Sci. (Chem. Sci.)* **1991**, *103* (5), 655–666. https://doi.org/10.1007/BF02841067.

(11) Santos, C. C.; Guedes, I.; Siqueira, J. P.; Misoguti, L.; Zilio, S. C.; Boatner, L. A. Third-Order Nonlinearity of $Er^{3+}$-Doped Lead Phosphate Glass. *Appl. Phys. B* **2010**, *99*, 559–563. https://doi.org/10.1007/s00340-010-3922-0.

(12) Singh, D. J.; Jellison, G. E.; Boatner, L. A. Electronic Structure of Pb- and Non-Pb-Based Phosphate Scintillators. *Phys. Rev. B* **2006**, *74*, 155126. https://doi.org/10.1103/PhysRevB.74.155126.

(13) Brixner, L. H.; Bierstedt, P. E.; Foris, C. M. Crystal Growth and Properties of Lead Pyrophosphate, $Pb_2P_2O_7$. *J. Solid State Chem.* **1973**, *6*, 430. https://doi.org/10.1016/0022-4596(73)90234-X.

(14) Schneider, M. Oriented Formation of $Pb_2P_2O_7$ from $K_2Pb[P_4O_{12}]$. *Z. Anorg. Allg. Chem.* **1983**, *503*, 238–240. https://doi.org/10.1002/zaac.19835030826.

(15) Desai, C. C.; Ramana, M. S. V. Crystal Data for $PbHPO_4$, $Pb_4(NO_3)_2(PO_4)_2{\cdot}2H_2O$ and $Pb_2P_2O_7$ Crystals. *J. Mater. Sci. Lett.* **1988**, *7*, 56–58. https://doi.org/10.1007/BF01729915.

(16) Mullica, D. F.; Perkins, H. O.; Grossie, D. A.; Boatner, L. A.; Sales, B. C. Structure of Dichromate-Type Lead Pyrophosphate, $Pb_2P_2O_7$. *J. Solid State Chem.* **1986**, *62*, 371–376. https://doi.org/10.1016/0022-4596(86)90252-5.

(17) Brückner, A.; Worza, H. Zur Kenntnis Einer Hochtemperaturmodifikation des Bleidiphosphats, $Pb_2P_2O_7$. *Z. Anorg. Allg. Chem.* **1990**, *584*, 173–177. https://doi.org/10.1002/zaac.19905840118.

(18) Elmarzouki, A.; Boukhari, A.; Holt, E. M.; Berrada, A. Study of $Sr_{2-x}Pb_xP_2O_7$ ($0 < x < 2$) Solid Solutions and Single-Crystal Structure of $Sr_{1.34}Pb_{0.66}P_2O_7$. *J. Alloys Compd.* **1994**, *204*, 127–131. https://doi.org/10.1016/0925-8388(94)90081-7.

(19) Webb, N. C. The Crystal Structure of β-$Ca_2P_2O_7$. *Acta Crystallogr.* **1966**, *21*, 942. https://doi.org/10.1107/S0365110X66004225.

(20) Boudin, S.; Grandin, A.; Borel, M. M.; Leclaire, A.; Raveau, B. Redetermination of the β-$Ca_2P_2O_7$ Structure. *Acta Crystallogr. C* **1993**, *49*, 2062–2064. https://doi.org/10.1107/S0108270193005608.

(21) Mbarek, A.; Edhokkar, F. The $P4_3$ Enantiomorph of $Sr_2As_2O_7$. *Acta Crystallogr. E* **2013**, *69*, i84. https://doi.org/10.1107/S1600536813031619.

(22) Edhokkar, F.; Hadrich, A.; Graia, M.; Mhiri, T. Synthesis and Crystal Structure of $Sr_2As_2O_7$ from Single-Crystal Data. *IOP Conf. Ser.: Mater. Sci. Eng.* **2012**, *28*, 012017. https://doi.org/10.1088/1757-899X/28/1/012017.

(23) Joung, M. R.; Kim, J. S.; Song, M. E.; Nahm, S. Formation Process and Microwave Dielectric Properties of $R_2V_2O_7$ (R = Ba, Sr, and Ca) Ceramics. *J. Am. Ceram. Soc.* **2009**, *92* (12), 3092–3094. https://doi.org/10.1111/j.1551-2916.2009.03324.x.

(24) Zhou, Y.; Kang, S. Z.; Qin, L.; Li, X. Boosting Charge Separation and Nitrogen Vacancies in Graphitic Carbon Nitride by Implanted Strontium Vanadate for Highly Efficient Photocatalytic Reduction of Hexavalent Chromium. *RSC Adv.* **2021**, *11*, 16034. https://doi.org/10.1039/D1RA01489G.

(25) Zhou, Z.; Wang, N.; Zhou, N.; He, Z.; Liu, S.; Liu, Y.; Tian, Z.; Mao, Z.; Hintzen, H. T. High Colour Purity Single-Phased Full Colour Emitting White LED Phosphor $Sr_2V_2O_7$:$Eu^{3+}$. *J. Phys. D: Appl. Phys.* **2013**, *46*, 035104. https://doi.org/10.1088/0022-3727/46/3/035104.

(26) Gupta, S. K.; Sudarshan, Y.; Kadam, R. M. Tunable White Light Emitting $Sr_2V_2O_7$:$Bi^{3+}$ Phosphors: Role of Bismuth Ion. *Mater. Des.* **2017**, *130*, 208–214. https://doi.org/10.1016/j.matdes.2017.05.056.

(27) Singh, R.; Dhoble, S. J. $Eu^{3+}$ and $Dy^{3+}$ Activated $Sr_2V_2O_7$ Phosphor for Solid State Lighting. *Adv. Mater. Lett.* **2011**, *2* (5), 341–344. https://doi.org/10.5185/amlett.2011.3071am2011.

(28) Christensen, N.; Hazell, R. G.; Hewat, A. W. Synthesis, Crystal Growth and Structure Investigations of Rare-Earth Disilicates and Rare-Earth Oxyapatites. *Acta Chem. Scand.* **1997**, *51*, 37–43. https://doi.org/10.3891/acta.chem.scand.51-0037.

(29) Halasyamani, P. S.; Poeppelmeier, K. R. Noncentrosymmetric Oxides. *Chem. Mater.* **1998**, *10*, 2753–2769. https://doi.org/10.1021/cm980140w.

(30) Feigelson, R. S. Synthesis and Single-Crystal Growth of Rare-Earth Orthophosphates. *J. Am. Ceram. Soc.* **1964**, *47*, 257. https://doi.org/10.1111/j.1151-2916.1964.tb14409.x.

(31) Smith, S. H.; Wanklyn, B. M. Flux Growth of Rare Earth Vanadates and Phosphates. *J. Cryst. Growth* **1974**, *21*, 23–28. https://doi.org/10.1016/0022-0248(74)90145-6.

(32) Rigaku Oxford Diffraction. *CrysAlisPro Software System*, version 1.171.41.119a; Rigaku Corporation: Oxford, 2021.

(33) Dolomanov, O. V.; Bourhis, L. J.; Gildea, R. J.; Howard, J. A. K.; Puschmann, H. OLEX2: A Complete Structure Solution, Refinement and Analysis Program. *J. Appl. Crystallogr.* **2009**, *42* (2), 339–341. https://doi.org/10.1107/S0021889808042726.

(34) Sheldrick, G. M. SHELXT—Integrated Space-Group and Crystal-Structure Determination. *Acta Crystallogr. A* **2015**, *71* (1), 3–8. https://doi.org/10.1107/S2053273314026370.

(35) Momma, K.; Izumi, F. VESTA 3 for Three-Dimensional Visualization of Crystal, Volumetric and Morphology Data. *J. Appl. Crystallogr.* **2011**, *44* (6), 1272–1276. https://doi.org/10.1107/S0021889811038970.

(36) Degen, T.; Sadki, M.; Bron, E.; König, U.; Nénert, G. The HighScore Suite. *Powder Diffr.* **2014**, *29* (S2), S13–S18. https://doi.org/10.1017/S0885715614000840.

(37) Hasabeldaim, E. H. H.; Swart, H. C.; Kroon, R. E. Luminescence and Stability of Tb-Doped $CaF_2$ Nanoparticles. *RSC Adv.* **2023**, *13*, 5353. https://doi.org/10.1039/d2ra07897j.

For Table of Contents Use Only,

Photoluminescent Tetragonal Tb-doped $Pb_2P_2O_7$

Yong Liu, Wenhua Bi, Alla Arakcheeva and Arnaud Magrez

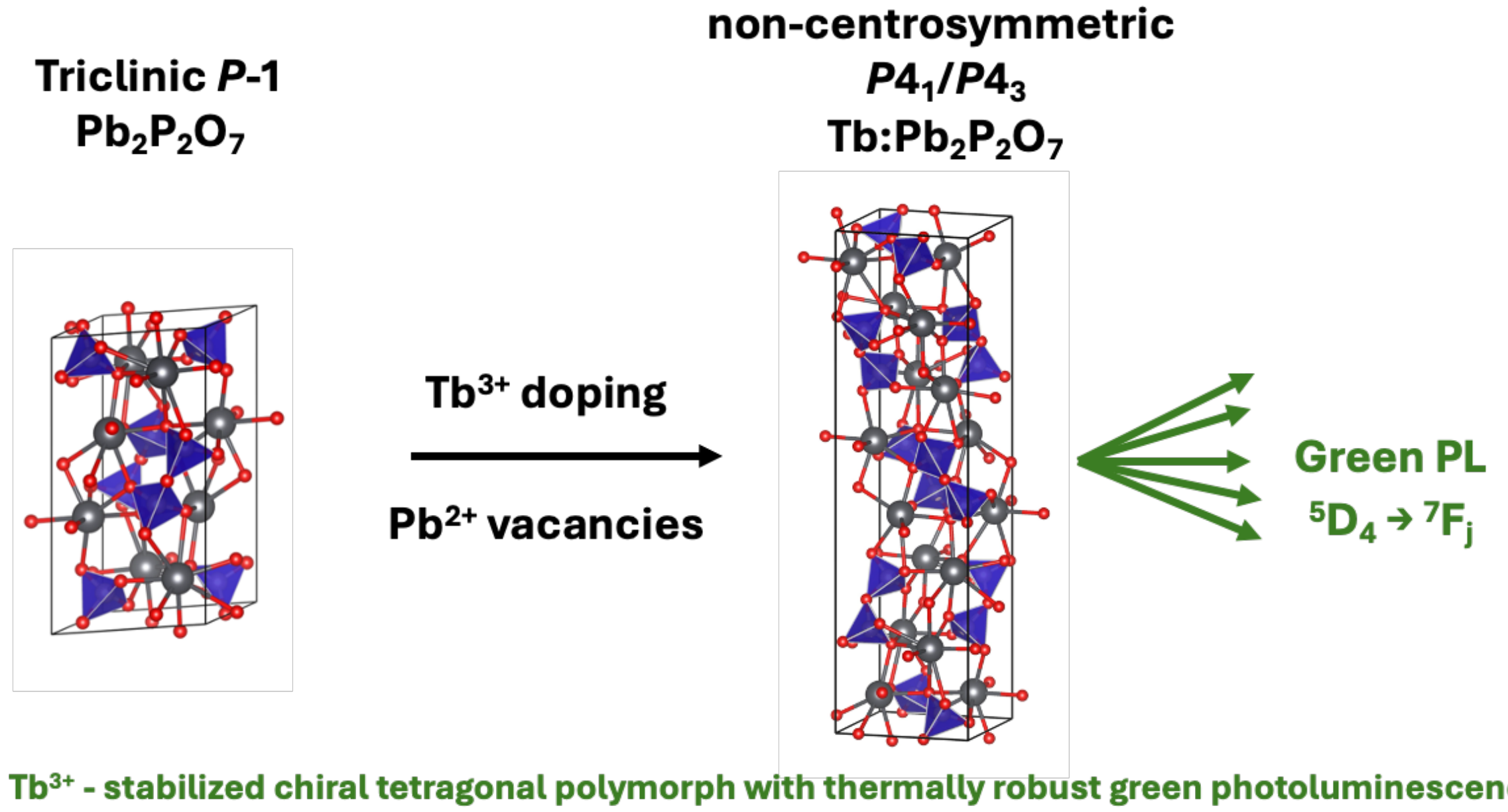

# Photoluminescent Tetragonal Tb-doped $Pb_2P_2O_7$

*Yong Liu, Wenhua Bi, Alla Arakcheeva and Arnaud Magrez*

Crystal Growth Facility, Institute of Physics, Ecole Polytechnique Fédérale de Lausanne, CH-1015 Lausanne

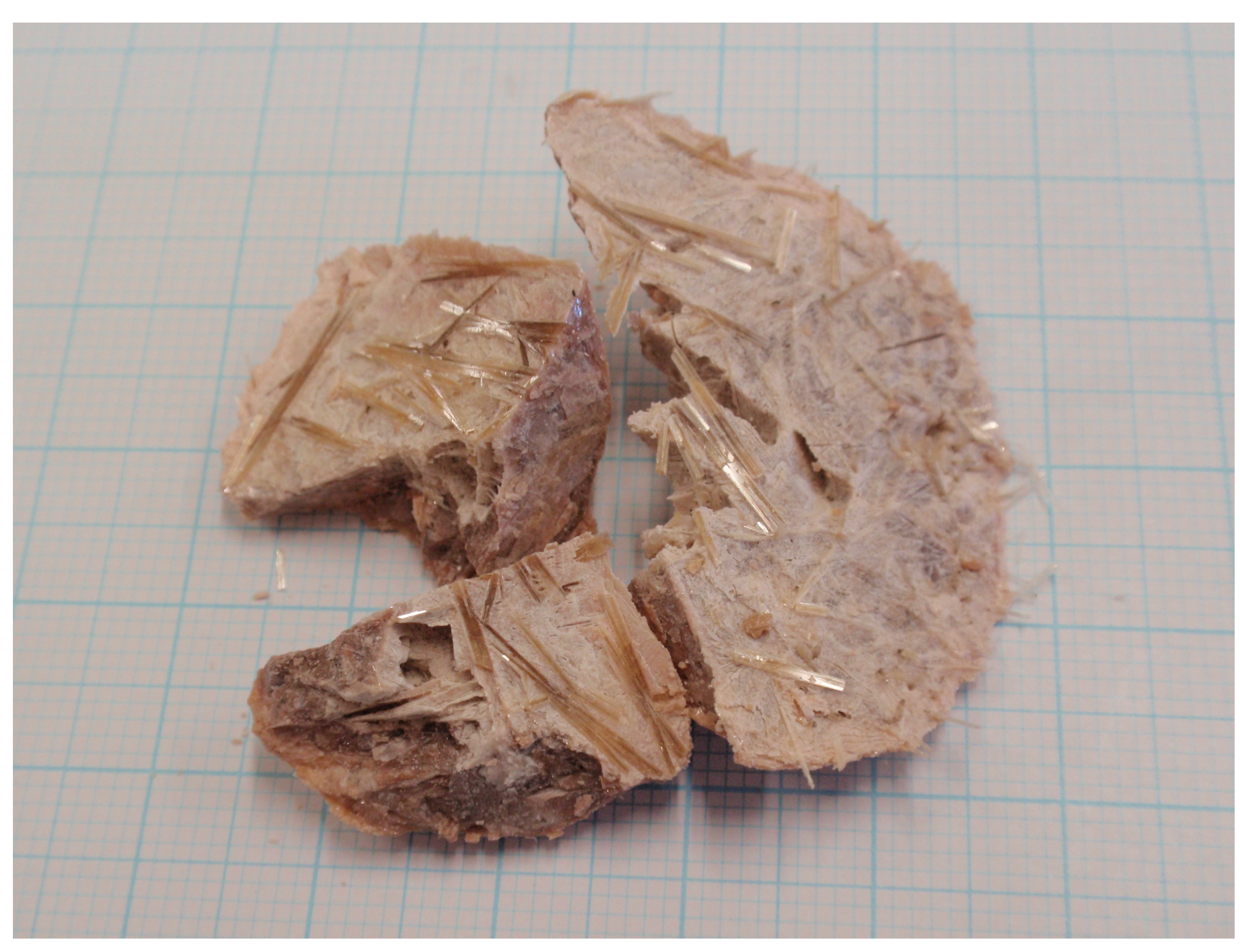

**Figure S1** Needle like $TbPO_4$ crystals embedded in a Tb doped $Pb_2P_2O_7$ flux.

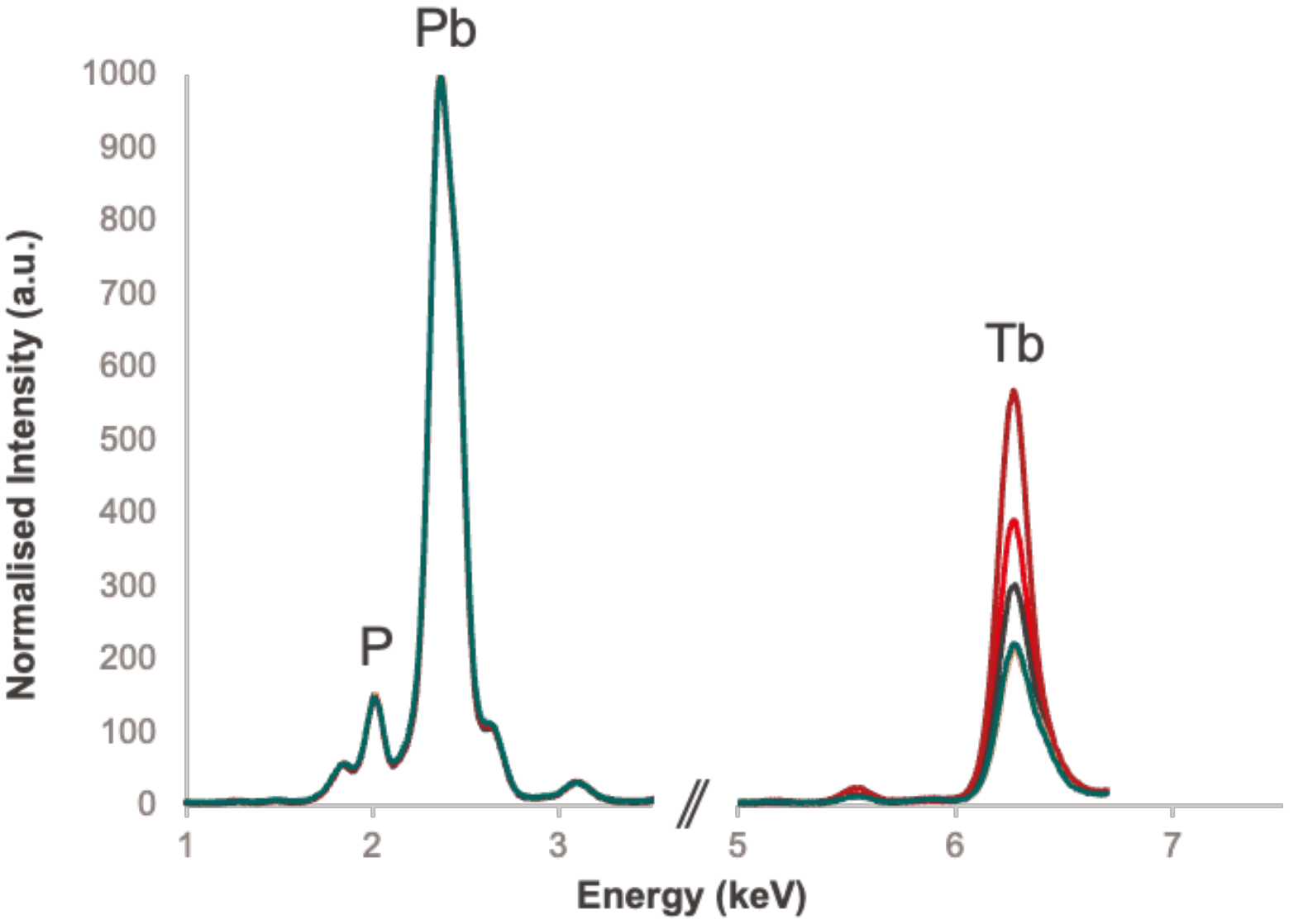

| Crystals | Tb/Pb |
|---|---|
| #1 | 2,5% |
| #2 | 2,7% |
| #3 | 1,9% |
| #4 | 1,6% |
| #5 | 1,3% |
| #6 | 3,7% |
| #7 | 1,1% |
| #8 | 1,4% |
| #9 | 1,3% |
| #10 | 1,9% |

**Figure S2**: XRF spectra of crystals with different Tb concentration. The composition of 10 different crystals are given in at%.

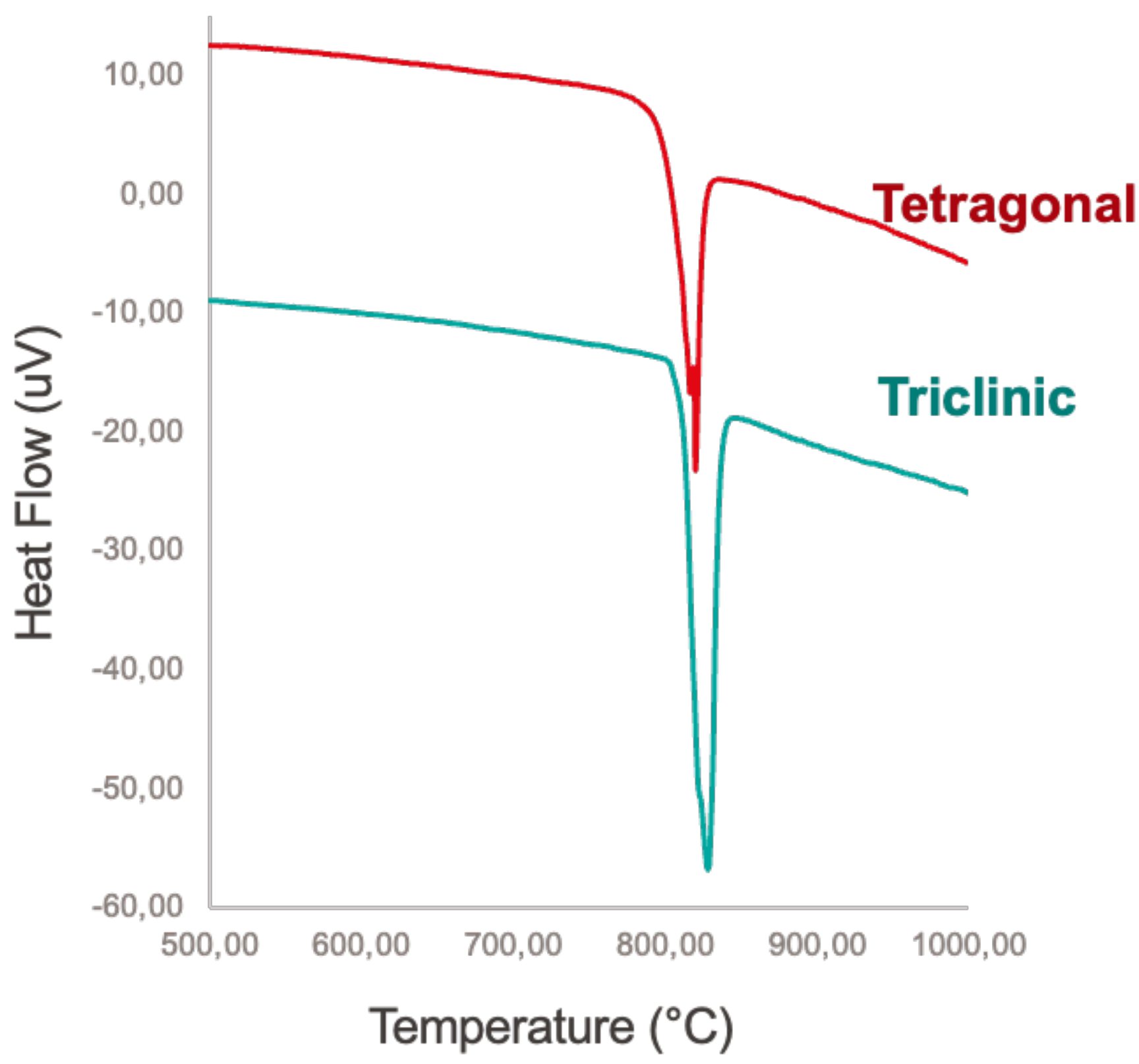

**Figure S3** DSC curves of Tb doped $Pb_2P_2O_7$ (red) and triclinic $Pb_2P_2O_7$ (blue). The endothermic peak at 800°C and 809°C respectively correspond to melting.

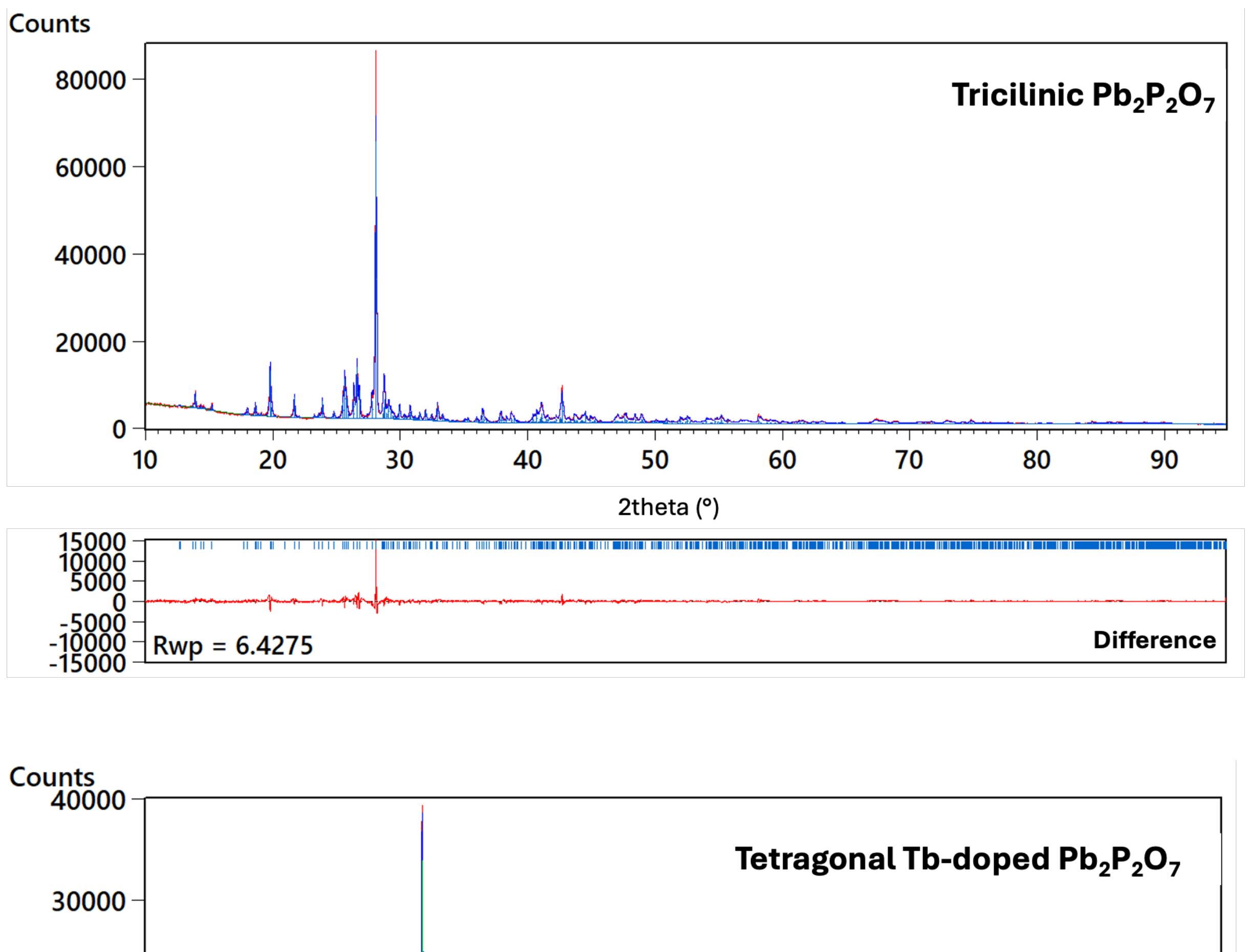

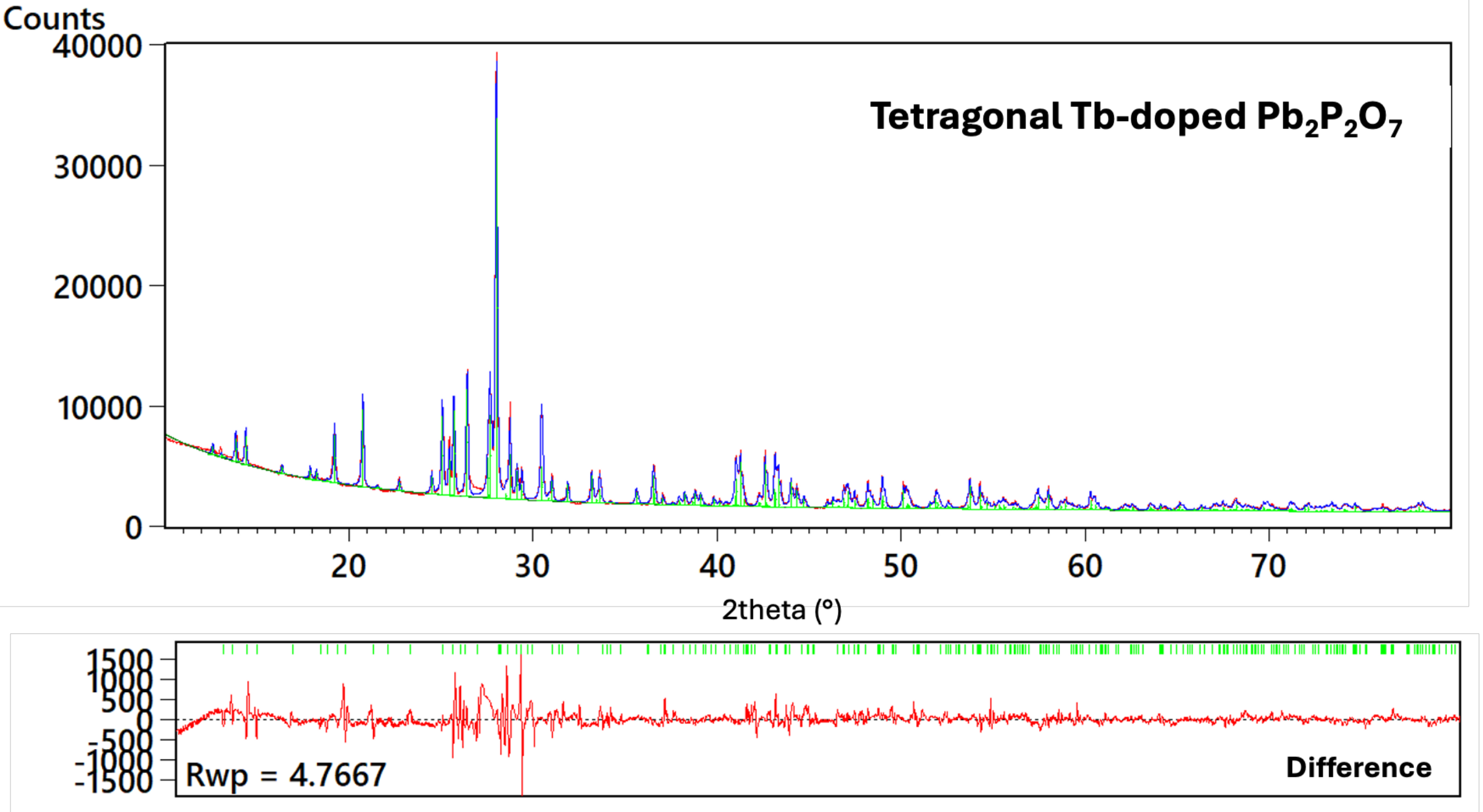

**Figure S4:** Profile fitting of the powder XRD pattern for the triclinic Pb2P2O7 (up) and tetragonal Tb-doped Pb2P2O7 (down). The difference between the calculated and experimental patterns is given in the panel below the patterns. The absence of extra-reflections confirm that both triclinic $Pb_2P_2O_7$ and tetragonal Tb-doped $Pb_2P_2O_7$ samples are single phase.

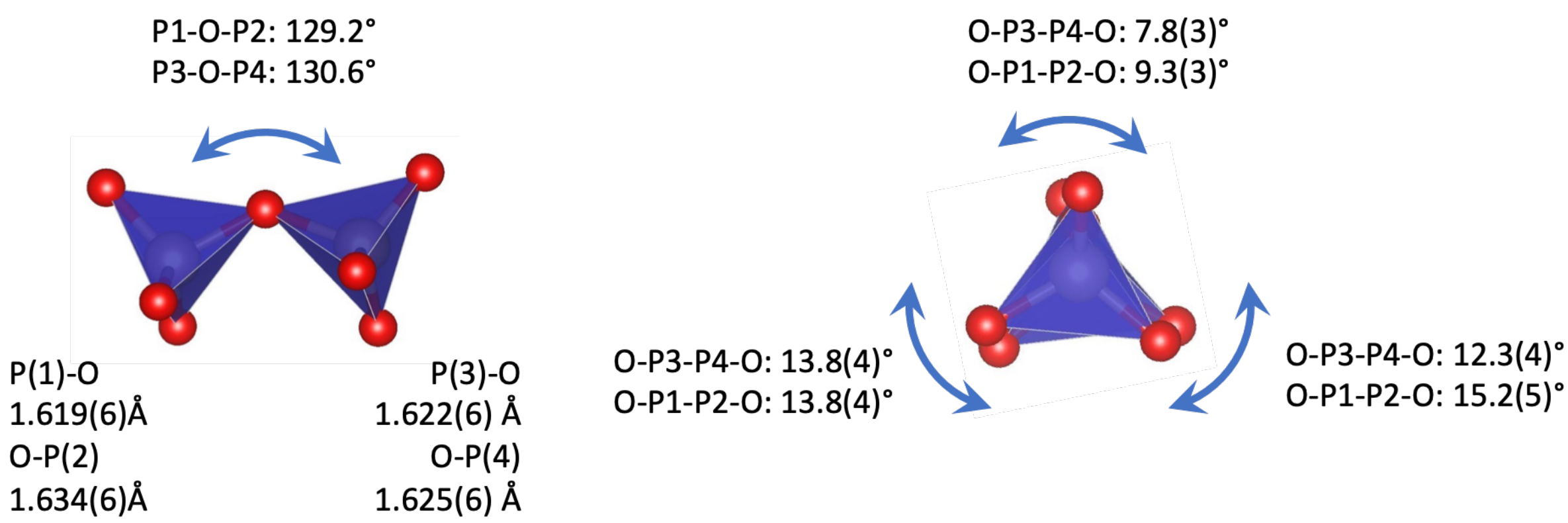

**Figure S5**: Pyrophosphates groups in the triclinic structure of $Pb_2P_2O_7$

### Table S1. Experimental details for tetragonal structure determination

| | I | II |
|---|---|---|
| **Crystal data** | | |
| Chemical formula | $(Pb_{1.989}Tb_{0.007})P_2O_7$ | $(Pb_{1.98}Tb_{0.012})P_2O_7$ |
| $M_r$ | 587.2 | 586.7 |
| Crystal system, space group | Tetragonal, $P4_1$ | Tetragonal, $P4_3$ |
| Temperature (°C) | 100K | 100K |
| *a*, *c* (Å) | 6.9577 (1), 25.3713 (6) | 6.9462 (1), 25.3661 (3) |
| *V* (Å$^3$) | 1228.21 (4) | 1223.91 (3) |
| *Z* | 8 | 8 |
| Radiation type | Mo *K*α | Mo *K*α |
| $\mu$ (mm$^{-1}$) | 55.05 | 55.19 |
| Crystal size (mm) | 0.16 × 0.12 × 0.06 | 0.11 × 0.08 × 0.07 |
| | | |
| **Data collection** | | |
| Diffractometer | XtaLAB Synergy, Single source at home/near, HyPix3000 | XtaLAB Synergy, Single source at home/near, HyPix3000 |
| Absorption correction | Gaussian, *CrysAlis PRO* 1.171.41.119a (Rigaku Oxford Diffraction, 2021) Numerical absorption correction based on gaussian integration over a multifaceted crystal model Empirical absorption correction using spherical harmonics, implemented in SCALE3 ABSPACK scaling algorithm. | Gaussian, *CrysAlis PRO* 1.171.41.119a (Rigaku Oxford Diffraction, 2021) Numerical absorption correction based on gaussian integration over a multifaceted crystal model Empirical absorption correction using spherical harmonics, implemented in SCALE3 ABSPACK scaling algorithm. |
| $T_{min}$, $T_{max}$ | 0.014, 0.158 | 0.058, 0.220 |
| No. of measured, independent and observed [$I > 3\sigma(I)$] reflections | 12240, 2630, 2274 | 32660, 5925, 5527 |
| $R_{int}$ | 0.082 | 0.048 |
| $(\sin \theta/\lambda)_{max}$ (Å$^{-1}$) | 0.649 | 0.836 |
| Range of *h*, *k*, *l* | $h = -8\rightarrow8$, $k = -8\rightarrow8$, $l = -31\rightarrow32$ | $h = -11\rightarrow11$, $k = -11\rightarrow11$ $l = -42\rightarrow42$ |
| | | |
| **Refinement** | | |
| $R[F^2 > 2\sigma(F^2)]$, $wR(F^2)$, $S$ | 0.036, 0.068, 0.97 | 0.025, 0.053, 1.21 |
| No. of reflections | 2630 | 5925 |
| No. of parameters | 197 | 200 |
| $\Delta\rho_{max}$, $\Delta\rho_{min}$ (e Å$^{-3}$) | 1.80, -1.53 | 2.08, -1.46 |
| Absolute structure | 1258 of Friedel pairs used in the refinement | 2896 of Friedel pairs used in the refinement |
| Absolute structure parameter | 0.196 (14) | -0.010 (6) |

Computer programs: *CrysAlis PRO* 1.171.41.119a (Rigaku OD, 2021), *SHELXT* 2018/2 (Sheldrick, 2018), Petricek, V., Dusek, M. & Palatinus L. (2014). Z. Kristallogr. 229 (5), 345-352., Brandenburg, K. & Putz, H. (2005). *DIAMOND* Version 3. Crystal Impact GbR, Postfach 1251, D-53002 Bonn, Germany.

## Table S2. Experimental details for triclinic structure determination

| | |
|---|---|
| Chemical formula | $Pb_2 P_2O_7$ (plus about 1%Tb in Pb position) |
| $M_r$ | 588.32 |
| Crystal system, space group | Triclinic, *P*-1 |
| Temperature (K) | 100 |
| *a*, *b*, *c* (Å) | 6.9455 (1), 6.9555 (1), 12.7621 (2) |
| *α*, *β*, *γ* (°) | 83.023 (1), 88.688 (1), 89.722 (1) |
| *V* (Å$^3$) | 611.80 (2) |
| *Z* | 4 |
| Radiation type | Mo *K*α |
| $\mu$ (mm$^{-1}$) | 55.05 |
| Crystal size (mm) | 0.13 × 0.11 × 0.09 |
| Data collection | |
| Diffractometer | XtaLAB Synergy, Single source at home/near, HyPix3000 |
| Absorption correction | Gaussian, *CrysAlis PRO* 1.171.41.119a (Rigaku Oxford Diffraction, 2021) Numerical absorption correction based on gaussian integration over a multifaceted crystal model Empirical absorption correction using spherical harmonics, implemented in SCALE3 ABSPACK scaling algorithm. |
| $T_{min}$, $T_{max}$ | 0.060, 0.209 |
| No. of measured, independent and observed [$I > 2\sigma(I)$] reflections | 18356, 3715, 3528 |
| $R_{int}$ | 0.047 |
| $(\sin \theta/\lambda)_{max}$ (Å$^{-1}$) | 0.714 |
| Refinement | |
| Range of *h*, *k*, *l* | $h = -9 \rightarrow 9$, $k = -9 \rightarrow 9$, $l = -18 \rightarrow 18$ |
| $R[F^2 > 2\sigma(F^2)]$, $wR(F^2)$, $S$ | 0.027, 0.067, 1.09 |
| No. of reflections | 3715 |
| No. of parameters | 200 |
| $\Delta\rho_{max}$, $\Delta\rho_{min}$ (e Å$^{-3}$) | 4.37, −2.52 |

Computer programs: *CrysAlis PRO* 1.171.41.119a (Rigaku OD, 2021), *CrysAlis PRO* 1.171.41.119a, *SHELXT2018*/3 (Sheldrick, 2018), *SHELXL2018*/3, Diamond, *publCIF*.

**Table S3. Atomic parameters for tetragonal structure of $(Pb_{1.98}Tb_{0.012})P_2O_7$ in $P4_3$ space group**

| Site occupancy* | x | y | z | Ueq |
|---|---|---|---|---|
| Pb1 | 0.16067(4) | 0.40324(3) | 0.432728(12) | 0.00552(5) |
| 0.985(3) Pb2 + 0.010(2) Tb2 | 0.97017(3) | -0.29914(3) | 0.573163 | 0.00565(6) |
| Pb3 | 0.38971(3) | 0.30292(3) | 0.572070(13) | 0.00574(5) |
| 0.980(3) Pb4 + 0.013(2) Tb4 | 0.73753(4) | 0.80526(3) | 0.427507(12) | 0.00531(6) |
| P1 | 0.8966(2) | 0.2252(2) | 0.54401(6) | 0.0048(4) |
| P2 | 0.6497(2) | 0.2831(2) | 0.45100(6) | 0.0045(4) |
| P3 | 0.4885(3) | 0.8166(2) | 0.55030(6) | 0.0051(4) |
| P4 | 0.2446(2) | 0.8640(2) | 0.45618(6) | 0.0050(4) |
| O1 | 0.9801(7) | 0.0477(7) | 0.56856(19) | 0.0096(12) |
| O2 | 1.0574(7) | 0.3594(7) | 0.52445(17) | 0.0060(11) |
| O3 | 0.7566(7) | 0.3406(7) | 0.57851(18) | 0.0060(10) |
| O4 | 0.7769(7) | 0.1564(7) | 0.49183(17) | 0.0066(11) |
| O5 | 0.6022(7) | 0.1428(7) | 0.40670(17) | 0.0068(11) |
| O6 | 0.4747(7) | 0.3553(7) | 0.48093(17) | 0.0080(11) |
| O7 | 0.7775(7) | 0.4479(7) | 0.43079(19) | 0.0076(11) |
| O8 | 0.3835(7) | 0.7475(7) | 0.42194(17) | 0.0071(11) |
| O9 | 0.1645(7) | 1.0389(7) | 0.42875(18) | 0.0079(11) |
| O10 | 0.0857(7) | 0.7309(7) | 0.47729(18) | 0.0073(11) |
| O11 | 0.3643(7) | 0.9397(7) | 0.50645(18) | 0.0067(11) |
| O12 | 0.5378(7) | 0.9664(7) | 0.59161(17) | 0.0074(11) |
| O13 | 0.6646(7) | 0.7372(7) | 0.52170(17) | 0.0077(11) |
| O14 | 0.3553(7) | 0.6619(7) | 0.5723(2) | 0.0079(11) |

*Wyckoff position is 4*a* (x,y,z) for all atomic sites.

**Table S4. Atomic parameters for tetragonal structure of $(Pb_{1.989}Tb_{0.007})P_2O_7$ in $P4_1$ space group**

| Site occupancy* | x | y | z | Ueq |
|---|---|---|---|---|
| Pb1 | 0.09815(9) | 0.34095(10) | 0.4342 | 0.01313(19) |
| Pb2 | 0.79878(9) | 0.53239(9) | 0.57488(4) | 0.01403(17) |
| Pb3 | 0.19577(9) | 1.11099(9) | 0.57355(4) | 0.01407(19) |
| 0.978(7)Pb4 + 0.01500(6) Tb4 | 0.69521(9) | -0.23584(10) | 0.42918(4) | 0.0132(2) |
| P1 | 0.2195(7) | 0.8509(7) | 0.45186(19) | 0.0133(12) |
| P2 | 0.2736(7) | 0.6046(7) | 0.54534(19) | 0.0117(12) |
| P3 | 0.6379(7) | 0.2590(7) | 0.4563(2) | 0.0137(12) |
| P4 | 0.6828(7) | 0.0161(7) | 0.55125(19) | 0.0117(12) |
| O1 | 0.1459(16) | 1.0273(16) | 0.4810(4) | 0.012(4) |
| O2 | 0.8368(16) | 0.1495(17) | 0.5733(6) | 0.019(4) |
| O3 | 0.7609(17) | -0.1609(18) | 0.5221(5) | 0.018(4) |
| O4 | 0.7519(16) | 0.1219(18) | 0.4233(5) | 0.016(4) |
| O5 | 0.1594(16) | 0.7434(15) | 0.5794(5) | 0.012(2) |
| O6 | 0.5313(17) | -0.0303(18) | 0.5921(4) | 0.016(4) |
| O7 | 0.4524(16) | 0.5232(17) | 0.5695(5) | 0.015(3) |
| O8 | 0.0516(16) | 0.7224(16) | 0.4315(5) | 0.014(3) |
| O9 | 0.4604(16) | 0.3363(16) | 0.4293(4) | 0.015(4) |
| O10 | 0.3583(16) | 0.8989(18) | 0.4082(4) | 0.014(4) |
| O11 | 0.3412(17) | 0.7243(17) | 0.4924(5) | 0.013(3) |
| O12 | 0.7677(17) | 0.4145(18) | 0.4782(5) | 0.017(3) |
| O13 | 0.1440(18) | 0.4467(15) | 0.5248(4) | 0.014(4) |
| O14 | 0.5593(16) | 0.1401(16) | 0.5075(5) | 0.012(3) |

*Wyckoff position is 4*a* (x,y,z) for all atomic sites.

**Table S5. Parameters of the triclinic $Pb_2P_2O_7$ structure.**

| | | | x | y | z | Occ. | U | Site |
|---|---|---|---|---|---|---|---|---|
| 1 | Pb | Pb1 | 0.34827(4) | 0.87579(4) | -0.36670(2) | 1.000 | 0.00545(7) | 2i |
| 2 | Pb | Pb2 | 0.91056(4) | -0.31225(4) | -0.14689(2) | 1.000 | 0.00563(7) | 2i |
| 3 | Pb | Pb3 | 0.32643(4) | 0.28861(4) | -0.14025(2) | 1.000 | 0.00570(7) | 2i |
| 4 | Pb | Pb4 | 0.75384(4) | 0.29435(4) | -0.34969(2) | 1.000 | 0.00550(7) | 2i |
| 5 | P | P1 | 0.8139(3) | 0.8007(3) | -0.41109(15) | 1.000 | 0.0053(3) | 2i |
| 6 | P | P2 | 0.7679(3) | 0.6024(3) | -0.59953(15) | 1.000 | 0.0050(3) | 2i |
| 7 | P | P3 | 0.8342(3) | 0.1981(3) | -0.08486(14) | 1.000 | 0.0043(3) | 2i |
| 8 | P | P4 | 0.5760(3) | 0.2080(3) | 0.10120(14) | 1.000 | 0.0052(3) | 2i |
| 9 | O | O1 | 0.6779(7) | 0.9696(8) | -0.4495(4) | 1.000 | 0.0066(9) | 2i |
| 10 | O | O2 | 0.6986(8) | 0.6440(8) | -0.3427(4) | 1.000 | 0.0077(10) | 2i |
| 11 | O | O3 | 0.9872(7) | 0.8690(8) | -0.3581(4) | 1.000 | 0.0083(10) | 2i |
| 12 | O | O4 | 0.8896(7) | 0.7111(8) | -0.5159(4) | 1.000 | 0.0067(9) | 2i |
| 13 | O | O5 | 0.6049(7) | 0.7389(8) | -0.6384(4) | 1.000 | 0.0075(10) | 2i |
| 14 | O | O6 | 0.6978(8) | 0.4131(8) | -0.5379(4) | 1.000 | 0.0076(10) | 2i |
| 15 | O | O7 | 0.9142(8) | 0.5772(8) | -0.6877(4) | 1.000 | 0.0076(10) | 2i |
| 16 | O | O8 | 0.6983(8) | 0.3322(8) | -0.1536(4) | 1.000 | 0.0075(10) | 2i |
| 17 | O | O9 | 0.9168(8) | 0.0338(8) | -0.1375(4) | 1.000 | 0.0078(10) | 2i |
| 18 | O | O10 | 0.9926(7) | 0.3202(7) | -0.0426(4) | 1.000 | 0.0059(9) | 2i |
| 19 | O | O11 | 0.7052(7) | 0.1049(7) | 0.0160(4) | 1.000 | 0.0059(9) | 2i |
| 20 | O | O12 | 0.5223(8) | 0.0419(8) | 0.1837(4) | 1.000 | 0.0070(9) | 2i |
| 21 | O | O13 | 0.7055(8) | 0.3542(7) | 0.1460(4) | 1.000 | 0.0064(9) | 2i |
| 22 | O | O14 | 0.4023(7) | 0.3006(8) | 0.0439(4) | 1.000 | 0.0066(9) | 2i |

**Table S6: Pb-O interatomic distances**

**Distances in Tb doped $Pb_2P_2O_7$ with $P4_1$ space group**

| (Pb1-O1) = 2.506(11) Å | (Pb2-O2) = 2.677(12) Å |
|---|---|
| (Pb1-O2) = 3.094(15) Å | (Pb2-O3) = 2.533(13) Å |
| (Pb1-O4) = 2.864(12) Å | (Pb2-O4) = 2.979(13) Å |
| (Pb1-O6) = 2.655(11) Å | (Pb2-O5) = 2.909(11) Å |
| (Pb1-O8) = 2.675(12) Å | (Pb2-O7) = 2.415(12) Å |
| (Pb1-O9) = 2.524(12) Å | (Pb2-O9) = 2.855(11) Å |
| (Pb1-O12) = 2.606(13) Å | (Pb2-O10) = 3.219(12) Å |
| (Pb1-O13) = 2.434(11) Å | (Pb2-O12) = 2.595(13) Å |
| | (Pb2-O13) = 2.782(12) Å |
| (Pb3-O1) = 2.444(11) Å | (Pb4/Tb4-O3) = 2.457(13) Å |
| (Pb3-O2) = 2.512(12) Å | (Pb4/Tb4-O4) = 2.525(13) Å |
| (Pb3-O5) = 2.574(11) Å | (Pb4/Tb4-O5) = 2.608(13) Å |
| (Pb3-O6) = 2.576(12) Å | (Pb4/Tb4-O8) = 2.497(12) Å |
| (Pb3-O8) = 2.828(13) Å | (Pb4/Tb4-O10) = 2.580(12) Å |
| (Pb3-O10) = 2.830(11) Å | (Pb4/Tb4-O11) = 2.952(13) Å |
| (Pb3-O13) = 2.667(11) Å | (Pb4/Tb4-O12) = 2.778(13) Å |
| (Pb3-O14) = 3.041(12) Å | |

**Distances in Tb doped $Pb_2P_2O_7$ with $P4_3$ space group**

| (Pb1-O2) = 2.454(5) Å | (Pb2/Tb2-O1) = 2.413(5) Å |
|---|---|
| (Pb1-O6) = 2.523(5) Å | (Pb2/Tb2-O2) = 2.742(5) Å |
| (Pb1-O9) = 2.533(5) Å | (Pb2/Tb2-O3) = 2.912(5) Å |
| (Pb1-O10) = 2.594(5) Å | (Pb2/Tb2-O5) = 3.219(5) Å |
| (Pb1-O12) = 2.645(5) Å | (Pb2/Tb2-O8) = 3.002(5) Å |
| (Pb1-O7) = 2.680(5) Å | (Pb2/Tb2-O9) = 2.877(5) Å |
| (Pb1-O8) = 2.862(5) Å | (Pb2/Tb2-O10) = 2.569(5) Å |
| (Pb1-O14) = 3.078(6) Å | (Pb2/Tb2-O13) = 2.505(5) Å |
| | (Pb2/Tb2-O14) = 2.689(5) Å |
| (Pb3-O6) = 2.414(5) Å | (Pb4/Tb4-O3) = 2.591(5) Å |
| (Pb3-O14) = 2.505(5) Å | (Pb4/Tb4-O4) = 2.947(5) Å |
| (Pb3-O3) = 2.567(5) Å | (Pb4/Tb4-O5) = 2.581(5) Å |
| (Pb3-O12) = 2.602(5) Å | (Pb4/Tb4-O7) = 2.499(5) Å |
| (Pb3-O2) = 2.635(5) Å | (Pb4/Tb4-O8) = 2.496(5) Å |
| (Pb3-O5) = 2.826(5) Å | (Pb4/Tb4-O10) = 2.777(5) Å |
| (Pb3-O7) = 2.843(5) Å | (Pb4/Tb4-O13) = 2.488(5) Å |
| (Pb3-O11) = 3.028(5) Å | |

**Distances in the triclinic $Pb_2P_2O_7$**

| (Pb1-O1) = 2.469(5) Å | (Pb4-O9) = 2.425(6) Å |
|---|---|
| (Pb1-O6) = 2.498(6) Å | (Pb4-O10) = 2.519(6) Å |
| (Pb1-O3) = 2.508(5) Å | (Pb4-O14) = 2.520(5) Å |
| (Pb1-O1) = 2.553(5) Å | (Pb4-O13) = 2.681(6) Å |
| (Pb1-O12) = 2.649(6) Å | (Pb4-O10) = 2.797(5) Å |
| (Pb1-O5) = 2.710(6) Å | (Pb4-O3) = 2.873(5) Å |
| (Pb1-O2) = 2.914(6) Å | (Pb4-O8) = 2.896(6) Å |
| (Pb1-O13) = 3.081(5) Å | (Pb4-O2) = 2.977(6) Å |
| | (Pb4-O7) = 3.184(6) Å |
| (Pb3-O14) = 2.432(6) Å | (Pb2-O2) = 2.472(6) Å |
| (Pb3-O13) = 2.486(5) Å | (Pb2-O6) = 2.482(6) Å |
| (Pb3-O8) = 2.601(6) Å | (Pb2-O5) = 2.513(5) Å |
| (Pb3-O10) = 2.628(5) Å | (Pb2-O7) = 2.553(6) Å |
| (Pb3-O12) = 2.637(6) Å | (Pb2-O8) = 2.568(6) Å |
| (Pb3-O7) = 2.853(6) Å | (Pb2-O1) = 2.783(6) Å |
| (Pb3-O5) = 2.883(6) Å | (Pb2-O4) = 2.983(5) Å |
| (Pb3-O11) = 2.998(5) Å | (Pb2-O9) = 3.290(6) Å |
| (Pb3-O9) = 3.357(6) Å | |